Banner appropriate to article type will appear here in typeset article

# On particle dynamics in steady axial rotor flows

**Francesco Caccia and Alberto Guardone**

Department of Aerospace Science and Technology, Politecnico di Milano
Via La Masa 34, 20156 Milan, Italy
**Corresponding author:** Alberto Guardone, alberto.guardone@polimi.it

We investigate the effect of rotor velocity induction on the distribution of particles impinging on rotor blades and model the delayed response of a particle to the rotor-induced velocity field. We consider as reference a wind turbine rotor and a small-scale propeller in axial flow conditions. We first show that the classical 2D modeling of the multi-phase flow can generate a systematic error with respect to the 3D solution. We consider two limiting cases: particles in equilibrium with the rotor-induced velocity field, where the carrier phase is computed using the section's aerodynamic velocity vector, and induction-independent particles, where the geometric velocity vector is used. The 3D solution differs from the two limiting cases when particles are in partial equilibrium with the induced velocity. We introduce an induction Stokes number $Stk_{ind}$ and identify a transition regime between the two limiting solutions for $0.1 \lesssim Stk_{ind} \lesssim 10$. Then, we present a simple 1D delay model to evaluate the induced component of the particle velocity at the rotor disk as a function $Stk_{ind}$. We validate the model by showing that it allows capturing the transition regime in 2D simulations. The model only requires knowledge of the aerodynamic and geometric velocity vectors, i.e., of the axial and tangential induction factors.

**Key words:** Authors should not enter keywords on the manuscript, as these must be chosen by the author during the online submission process and will then be added during the typesetting process (see Keyword PDF for the full list). Other classifications will be added at the same time.

## 1. Introduction

The interaction between dispersed particles and rotor flows represents a fundamental problem in fluid mechanics, central to applications ranging from energy conversion to atmospheric and environmental processes. From turbomachinery and propulsion devices to energy systems, rotors frequently operate in conditions where a dispersed phase interacts with the surrounding air. Examples include aircraft propellers, helicopter rotors, and wind turbines, as well as industrial fans and cooling towers. Such interactions are now encountered even in extraterrestrial atmospheres: between 2021 and 2024, the Ingenuity Mars Helicopter (Balaram *et al.* 2021) completed 72 flights in the dusty environment of Mars, marking a new era of planetary exploration. In all these contexts, suspended solid or liquid particles interact with the rotor blades and the induced flow field, whether intentionally or incidentally, making it essential to understand the underlying particle–flow dynamics.

Two illustrative examples of rotor–particle interaction are ice accretion and particle-induced erosion on rotor blades. These opposing yet physically related processes are governed by the interaction of particles with a solid surface, one leading to material accumulation and the other to material removal, mainly on the leading edge of the blade. Ice accretion occurs when supercooled water droplets impact a surface, breaking their unstable equilibrium and solidifying on the blade (Guardone *et al.* 2025). It occurs over a time frame ranging from minutes to hours, depending on rotor size and operating conditions, and the consequences span from aerodynamic degradation to safety hazard. In a longer time frame, the impact of sand and rain on blades may lead to leading-edge erosion, causing a permanent degradation of blade aerodynamics; interestingly, the study of drop impact dynamics originated to understand rain-induced erosion (Cheng *et al.* 2022). The problems are shared among rotors of different sizes and purposes, ranging from micro-scale propellers to helicopter and large wind turbine rotors.

The investigation of such problems on rotors has been pursued extensively through both experimental and numerical means. Special attention has been devoted to ice accretion due to its inherent safety risks. In the case of propellers, their relatively small size allows for full- or quasi-full-scale experimental testing in icing wind tunnels, as demonstrated in several campaigns, e.g., by Karli *et al.* (2024) and Yan *et al.* (2024), Hardenberg *et al.* (2024), and Müller & Hann (2022). Conversely, icing experiments on full-scale helicopters (Shaw & Ritcher 1985) are generally limited to natural icing conditions due to the constraints imposed by rotor size and test facility capabilities. Wind tunnel experiments are typically conducted on scaled models under controlled conditions (Britton 1994; Kind *et al.* 1998). For wind turbines, experimental studies involving multiphase flow are commonly restricted to scaled fixed (Hochart *et al.* 2008) or rotating (Han *et al.* 2012) sectional models due to the large scale of the system.

In the past, full three-dimensional numerical simulations remained limited due to a combination of lacking computational tools and high computational cost. Recent advancements in software development and high-performance computing have enabled the development of high-fidelity numerical approaches that offer a cost-effective and safer alternative to in-flight testing. Yet, reduced-order models based on isolated two-dimensional blade sections, originally developed for helicopter icing (Gent & Cansdale 1985), are still widely used in various modified forms, to reduce computational time, explore broader design spaces (Gallia *et al.* 2024) and complex events (Switchenko *et al.* 2014), or introduce relevant physics such as blade plunging and pitching motions (Narducci *et al.* 2012; Kelly *et al.* 2018; Min & Yee 2023) and the history of the angle of attack of a blade section operating in a turbulent boundary layer (Gimenez *et al.* 2024; Tahani *et al.* 2024).

In 2D approximations, the sectional flow field is generally computed considering the local angle of attack of the section, i.e., the angle between the local chord and the vector resulting from the composition of the freestream wind vector, the induced velocity vector, and the local section velocity, projected in the section plane. The angle is either chosen to be representative of normal operating conditions or estimated with some inexpensive aerodynamic model, e.g., the blade element momentum theory. At times, the induced velocity vector is even neglected for convenience (Heramarwan *et al.* 2023). The dispersed phase is then solved considering a one-way coupling and initial equilibrium with the flow field. The approach is shared for studying both ice accretion (Martini *et al.* 2022; Caccia & Guardone 2023; Caccia *et al.* 2024) and erosion (Castorrini *et al.* 2019).

To introduce the focus of this work, it is useful to analyse the works by Fiore & Selig for damage and erosion of wind turbine sections. In Fiore & Selig (2015), they investigated sand and insect damage considering the standard workflow described earlier, i.e., computing the sectional flow field at standard operating angles of attack and presumably introducing the dispersed phase in equilibrium with the section velocity vector. Later, Fiore *et al.* (2016) investigated rain and hailstone damage in similar sectional conditions. Since heavy particles were considered, these were injected in the sectional flow field with their terminal velocity as seen by the section, i.e., in non-equilibrium with the free-stream flow field seen by the section, which includes the induced velocity.

Let us provide a visual representation of the problem in figure 1. The diagrams illustrate the behaviour of two particles of different sizes as they approach a wind turbine blade section. The section is located at a distance $r$ from the rotation axis as highlighted in figure 1a, together with the freestream velocity vector $\boldsymbol{V}_\infty$, the rotation speed $\Omega$ and the tangential velocity $\Omega r$. We start considering a 3D description of the problem in figure 1b. The velocity triangle of the carrying phase results from the sum of $V_\infty$ with the axial induced velocity $aV_\infty$ (where $a$ is the axial induction factor) in the $x$ direction and of $\Omega r$ with the tangential induced velocity $a'\Omega r$ (where $a'$ is the tangential induction factor) in the $y$ direction. On a rotor providing momentum to the surrounding fluid, the direction of the axial and tangential induced velocities would be opposite.

Let us now introduce particles in the flow field of figure 1b. The behaviour of a particle in the flow field is determined by its Stokes number $Stk = \tau_p/\tau_f$, where $\tau_p$ is the particle relaxation time and $\tau_f$ is a characteristic timescale of the carrier flow. A sufficiently small particle, i.e., a particle with $Stk \ll 1$, will adapt almost instantly to the flow-field, leading to a *tracking trajectory* that closely follows the streamlines of the carrying phase. Larger particles require a longer time to adapt to the flow field; thus, they deviate from it, especially in high streamline curvature regions. If a particle coming from the freestream is sufficiently large ($Stk \gg 1$), it is unaffected by the flow features, including the induced velocity field, and follows a *ballistic trajectory*.

We then restrict the problem to an infinite wing, equivalent to a 2D sectional description. In this case, if we prescribe the angle of attack $\alpha_{\text{aero}}$ obtained considering the induced velocities in the freestream conditions seen by the section (figure 1c), the flow field close to the section will be almost equivalent to the three-dimensional one of figure 1b, in the absence of relevant 3D phenomena. Thus, a particle with $Stk \ll 1$ will follow the correct tracking trajectory. However, a particle with $Stk \gg 1$ released in equilibrium at the freestream will necessarily follow a wrong ballistic trajectory, which includes the induced velocity component. We note that this is the state-of-the-art description of the sectional problem, as presented, e.g., by Barfknecht & von Terzi (2024).

Conversely, if we do not consider the induced velocities for the freestream conditions seen by the section (figure 1d), we obtain a geometric angle of attack $\alpha_{\text{geom}}$. In this case, neither the carrier velocity field around the section, nor the tracking trajectories ($Stk \ll 1$) will represent the 3D solution. However, ballistic trajectories ($Stk \gg 1$) will be correct.

This informal analysis suggests that restricting the problem to a 2D analysis would likely result in two classes of correct solutions for the dispersed phase dynamics around the section, which depend on the Stokes number. For $Stk \ll 1$, the correct solution is obtained by matching the flow field, i.e., by using $\alpha_{\text{aero}}$ (figure 1c). We refer to this modelling approach as *2D Ind*. For $Stk \gg 1$, the correct solution is obtained by neglecting the induced velocity, i.e., by using $\alpha_{\text{geom}}$ (figure 1d). We name this modelling approach *2D Geom*.

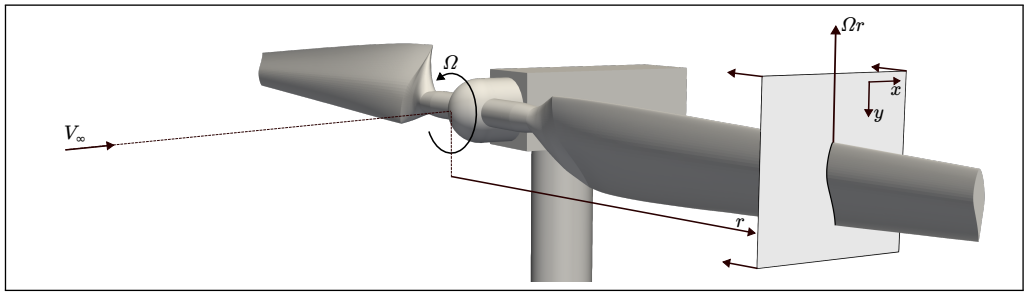

(a)

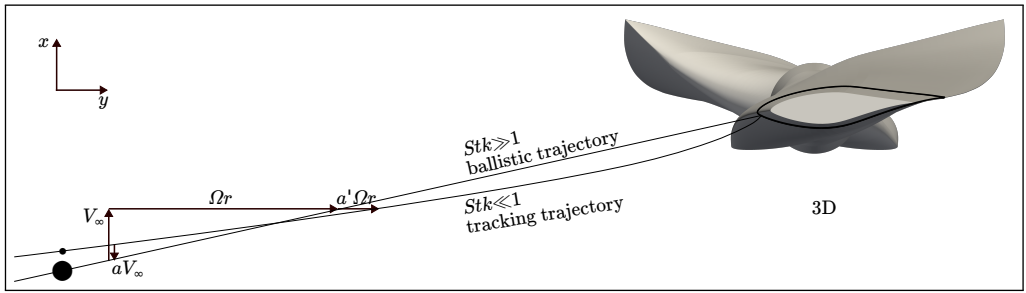

(b)

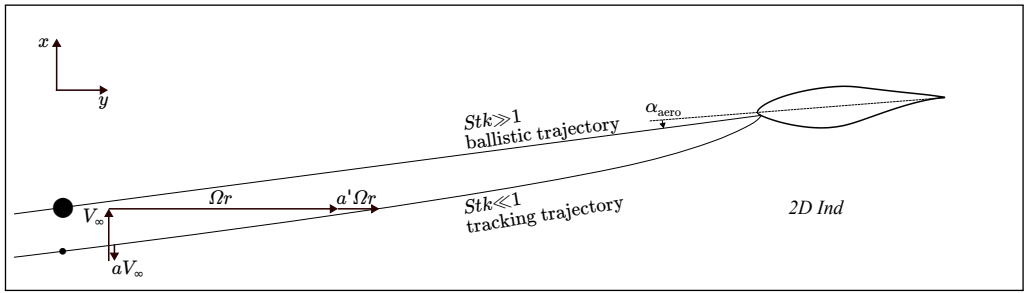

(c)

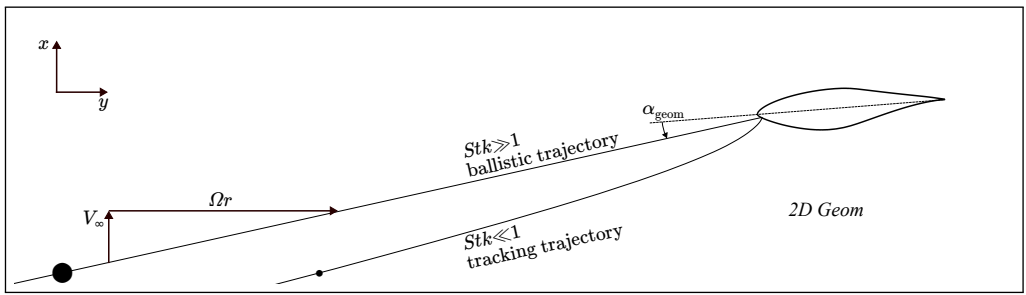

(d)

Figure 1. Representation of a tracking and a ballistic trajectory. (a) 3D global view of a wind turbine rotor, highlighting the plane analysed; (b) 3D solution in a reference frame fixed with the rotor: the tracking trajectory is affected by the induced velocity, and the ballistic trajectory is not; (c) *2D Ind* sectional description: the induced velocity contributes to the freestream velocity vector; (d) *2D Geom* sectional description: the induced velocity does not contribute to the freestream velocity vector.

In this work, we investigate the behaviour of particles impinging on rotor blades and their response to the induced velocity field through the comparison of the 3D solution with the *2D Ind* and *2D Geom* approaches. Through this comparison, we demonstrate that these are the two limiting cases and identify the range of Stokes numbers under which they provide the correct representation of the problem. Then, we present and validate an analytical model to compute the delayed response of a particle to the rotor-induced velocity field, which can be used in 2D sectional representations of the problem. The study is conducted numerically on a wind turbine rotor and a small-scale propeller in axial flow conditions, considering particle sizes relevant in real applications.

The paper is structured as follows. We start by defining the modelling assumptions and the numerical methods in § 2. In § 3 we present and discuss the results related to the 2D limiting solutions introduced in figure 1 through the comparison with 3D results. In § 4 we present and validate the model for computing the delayed response of a particle to the rotor-induced velocity field. At the end of the section, we summarise the transport regimes identified for the particles and the relationship between the observed phenomena and rotor size. Conclusions are drawn in § 6.

## 2. Methodology

We consider water droplets for the dispersed phase due to their relevance for both ice accretion and erosion events. The analysis is carried out under the following assumptions:

1. the carrier phase is described by the steady Reynolds-Averaged Navier-Stokes (RANS) equations. Only axial flight conditions with steady and uniform inflow velocity are considered, leading to a steady 3D solution when computed in a rotating reference frame. Also, the gravitational force acting on the particles is neglected, as it would introduce a time-dependent solution if the force is not aligned with the rotation axis;

2. the flow is modelled under a one-way coupling assumption, and the dispersed phase is influenced by the carrier phase, but not vice versa. This assumption is justified under a low volume fraction and mass loading of the dispersed phase (Balachandar & Eaton 2010), and is standard for the class of problems considered in this study;

3. particles stick upon impact and do not break up or rebound. Although this is an approximation for large water droplets, the assumption does not interfere with the phenomena observed in this work and allows generalisation to solid particles. Moreover, it will be possible to discuss the behaviour of secondary droplets in 2D and 3D simulations in view of the results obtained. Yarin (2006) reviewed such particle-surface interactions in detail.

Moreover, mono-dispersed clouds are studied, and each particle size is characterised independently. In practical applications, particle populations are typically polydisperse and their overall behaviour can be modelled as the superposition of multiple discrete particle-diameter bins.

The solution procedure is the same for both 2D and 3D simulations and consists of solving the dispersed phase after the carrying phase, as detailed in the following subsections. However, 2D simulations usually require an additional step to get the local boundary conditions on the section. Besides using generic operational angles of attack of the sections, common practices include using the Blade Element Momentum Theory (BEMT) to obtain the induced velocity at the blade section considering the operating conditions of the rotor, as reviewed by Martini *et al.* (2022) for wind turbine applications, and changing the angle of attack of the 2D simulation to match the pressure coefficient distribution on the pressure side of the corresponding section of the 3D rotor, as proposed by Narducci & Kreeger (2012) and Narducci *et al.* (2012) and recently applied by Oztekin & Bain (2024) on a propeller. To compute $\alpha_{\text{aero}}$ in the *2D Ind* case of figure 1c, we will cover both approaches, and use the former for the wind turbine rotor and the latter for the small-scale propeller. For the *2D Geom* case of figure 1d, $\alpha_{\text{geom}}$ is retrieved analytically.

### 2.1. *Solution of the carrying phase*

The flow field is computed by solving Reynolds-Averaged Navier-Stokes (RANS) equations in the open-source toolkit SU2 (Economon *et al.* 2015). SU2 solves the governing equations using a node-centered finite volume method on unstructured grids. In particular, the governing equations are written with an Arbitrary Lagrangian-Eulerian (ALE) differential form as

$$\frac{\partial \boldsymbol{U}}{\partial t} + \nabla \cdot \boldsymbol{F}^c_{\text{ALE}} - \nabla \cdot \boldsymbol{F}^v - \boldsymbol{Q} = 0 \,, \tag{2.1}$$

where $\boldsymbol{U} = \{\rho_f, \ \rho_f \boldsymbol{v}_f, \ \rho_f E\}^{\mathrm{T}}$ is the vector of conservative variables, $\boldsymbol{F}^c_{\text{ALE}}$ are the convective fluxes in the ALE formulation, $\boldsymbol{F}^v$ are the viscous fluxes and $\boldsymbol{Q}$ is a generic source term; $\rho_f$ is the fluid density, $\boldsymbol{v}_f$ is the fluid velocity in the inertial reference frame, and $E$ is the total energy per unit mass. The vector of convective fluxes $\boldsymbol{F}^c_{\text{ALE}}$ is

$$\boldsymbol{F}^c_{\text{ALE}} = \left\{ \begin{array}{c} \rho_f(\boldsymbol{v}_f - \boldsymbol{v}_\Omega) \\ \rho_f \boldsymbol{v}_f \otimes (\boldsymbol{v}_f - \boldsymbol{v}_\Omega) + \boldsymbol{I}p \\ \rho_f(\boldsymbol{v}_f - \boldsymbol{v}_\Omega)E + p\boldsymbol{v}_f \end{array} \right\} \,, \tag{2.2}$$

where $\boldsymbol{v}_\Omega$ is the velocity of the domain, $p$ is the static pressure, and $\boldsymbol{I}$ is the identity matrix. Let us consider a reference frame rotating with velocity $\boldsymbol{\Omega}$ and denote with $\boldsymbol{r}$ the distance vector from the rotation centre. By setting $\frac{\partial \boldsymbol{U}}{\partial t} = \boldsymbol{0}$, $\boldsymbol{v}_\Omega = \boldsymbol{\Omega} \times \boldsymbol{r}$, and $\boldsymbol{Q} = \{0, -\rho_f(\boldsymbol{\Omega} \times \boldsymbol{v}_f), 0\}^{\mathrm{T}}$, and by providing an appropriate turbulence closure for $\boldsymbol{F}^v$, one obtains the steady RANS equations in a non-inertial, rotating reference frame written in the absolute velocity formulation. In the RANS framework, closure is provided by the SA turbulence model from Spalart & Allmaras (1992), whereas we modelled laminar-turbulent transition with the algebraic BCM model by Cakmakcioglu *et al.* (2020).

In 2D, once the local boundary conditions are known and the grid is generated, the flow field is simulated on the sections chosen for the analysis. We discretized the convective fluxes with the Roe scheme, using a MUSCL scheme for flux reconstruction to recover second-order accuracy in space. Convergence was obtained after the reduction of the root mean square of the residual on the density by 6 orders of magnitude and the satisfaction of a Cauchy criterion for lift and drag over the last 200 iterations with a threshold value of $10^{-6}$. In 3D, the convective fluxes are discretized using the Jameson-Schmidt-Turkel (JST) scheme. Convergence is checked both on the root mean square of the residual of the density, where we aim at a reduction by 4 orders of magnitude and on the stabilisation of rotor thrust and torque with a Cauchy criterion as for the 2D simulations. In both cases, viscous fluxes are discretised with a first-order scalar upwind scheme. A free-stream turbulence intensity of 0.2% was considered.

### 2.2. *Solution of the discrete phase*

We use a discrete parcel method to track the position of computational parcels in the flow field, using the in-house Lagrangian particle tracking code PoliDrop (Bellosta *et al.* 2023). The trajectory of a particle of mass $m_p$, diameter $d_p$, and moving with velocity $\boldsymbol{v}_p$ is computed by integrating in time $t$ the following differential equation, where the only external force is the aerodynamic drag

$$\frac{\partial \boldsymbol{v}_p}{\partial t} = \frac{1}{m_p} \frac{\pi}{8} \left( \boldsymbol{v}_f - \boldsymbol{v}_p \right) \mu_f \, d_p \, Re_p \, C_{D,p} \, ,$$  (2.3)

where $\mu_f$ is the carrier phase dynamic viscosity, whereas $Re_p$ and $C_{D,p}$ are the particle's Reynolds number and drag coefficient, respectively. The former is defined as

$$Re_p = \frac{\rho_f \, ||\boldsymbol{v}_f - \boldsymbol{v}_p||_2 \, d_p}{\mu_f} \, ,$$  (2.4)

where $\rho_f$ is the carrier phase density. The drag coefficient of the droplet accounts for its deformation from a perfect sphere to an oblate disk due to the aerodynamic forces. It is expressed as a combination of the drag of a sphere $C_{D,\text{sphere}}$ and that of a disk $C_{D,\text{disk}}$ weighted on the eccentricity $\varepsilon$ and it is defined as

$$C_{D,p} = \begin{cases} (1-\varepsilon) \, C_{D,\text{sphere}} + \varepsilon \, C_{D,\text{disk}} & \text{if } We_p \leq 12 \\ C_{D,\text{disk}} & \text{if } We_p > 12 \end{cases}$$  (2.5)

where $\varepsilon = \left( 1 + 0.07\sqrt{We_p} \right)^{-6}$ depends on the Weber number of the particle $We_p = \rho_p ||\boldsymbol{v}_f - \boldsymbol{v}_p||^2 d_p / \sigma_p$, which expresses the ratio between inertia and surface tension forces; $\sigma_p$ is the droplet surface tension. For more details, the reader is referred to Bellosta *et al.* (2023), where the routines for parcel localisation and the solution interpolation are verified and the code is validated against experimental measurements on fixed wings.

A single, straight (2D) or planar (3D), uniform front of droplets is released in local equilibrium with the unperturbed flow. In 3D, the discrete phase is tracked in a rotating reference frame using an algorithm that avoids integrating the non-inertial accelerations, thus eliminating the discretisation error related to their integration. Details and verification are provided in Appendix A. After the particles' impact, an automatic cloud adaptation strategy is deployed to refine the initial cloud front and iteratively obtain a converged solution. The simulation is considered converged when the $\ell^2$ norm of the difference of the collection efficiency in two consecutive iterations is below a specified threshold. The threshold is set to $10^{-6}$ and $2 \times 10^{-3}$ for 2D and 3D simulations, respectively.

### 2.3. *Collection efficiency*

The parameter of interest in this study is the fraction of the dispersed phase freestream mass flux impacting at a given location on the surface. Depending on the application, it is referred to by various names, including *collection efficiency*, *collision efficiency*, *impingement efficiency*, and *hit rate*. In this work, we adopt the term *collection efficiency* and denote it by the symbol $\beta$. Numerically, the collection efficiency $\beta_i$ on the cell $i$ of the boundary is computed as

$$\beta_i = \frac{\sum\limits_{j \in i} m_{p,j} \phi_j}{A_i \rho_c^*} \, ,$$  (2.6)

where $A_i$ is the area of the cell at impact, $\rho_c^*$ is the computational cloud density computed as the ratio between the total cloud mass and the area of the injection front, and $m_j$ is the mass of the $j^{th}$ particle impinging on cell $i$. $\phi_j$ is a flux coefficient to account for the orientation of the cloud front at the freestream with respect to the relative velocity

$$\phi_j = \frac{V_{fs} - \boldsymbol{\Omega} \times \boldsymbol{r}_{0p,j}}{V_{fs}} \cdot \hat{\boldsymbol{n}}_c,$$ (2.7)

where $\hat{\boldsymbol{n}}_c$ is the unit vector normal to the seeding plane and $\boldsymbol{r}_{0p,j}$ is the $j^{th}$ particle distance vector from the rotation axis at the release location; $\boldsymbol{V}_{fs}$ represents the freestream velocity vector, and $V_{fs} = ||\boldsymbol{V}_{fs}||_2$ (this convention will be used from now on). In 3D, $\boldsymbol{V}_{fs} = \boldsymbol{V}_\infty$; in 2D, $\boldsymbol{V}_{fs}$ is the freestream velocity vector as seen by the section. This allows results independent of the orientation of the seeding plane in both the inertial and rotating reference frames. Then, the mass per unit time impinging on the cell can be computed as

$$\dot{m}_i = \rho_c \, V_{fs} \, A_i \, \beta_i \,,$$ (2.8)

where $\rho_c$ is the cloud density indicating the mass of dispersed phase contained in a unit volume of the carrying phase. In the case of water droplets, $\rho_c$ is generally named liquid water content (LWC). Finally, the local volumetric flux $\dot{\mu}$ and a section-normalized collection efficiency $\hat{\beta}$ are computed as

$$\dot{\mu}_i = \frac{\dot{m}_i}{\rho_c} \frac{1}{A_i}$$ (2.9)

and

$$\hat{\beta}_i = \frac{\dot{\mu}_i}{||\boldsymbol{V}_{fs} - \boldsymbol{\Omega} \times \boldsymbol{r}_i||_2} \,,$$ (2.10)

where $\boldsymbol{r}_i$ is the distance vector of the cell from the rotation center. In 3D, the first quantity shows the variation of impinging particle volume along the blade span, while the second provides a comparison with the fixed-wing equivalent collection efficiency. In 2D, $\hat{\beta}_i = \beta_i$, and $\hat{\beta}_i$ is therefore used for comparing 2D and 3D simulations consistently.

Finally, it is also possible to compute analytically the limit collection efficiency for an ideal ballistic trajectory as

$$\beta_{i,ball} = -\frac{V_{fs} - \boldsymbol{\Omega} \times \boldsymbol{r}_i}{V_{fs}} \cdot \hat{\boldsymbol{n}}_i \,,$$ (2.11)

where $\hat{\boldsymbol{n}}_i$ is the local surface normal vector, positive outwards; the result can be normalised according to equations (2.8) to (2.10). It is also necessary to set $\beta_{i,ball} = 0$ on surface cells that are in the shadow of other cells, i.e., only the first crossing of the ballistic trajectory with the body should be considered; this automatically excludes any cell with $\beta_{i,ball} < 0$.

### 2.4. *Stokes Numbers*

The results are presented and discussed in terms of the Stokes number $Stk = \tau_p/\tau_f$. In the Stokesian regime, i.e., if the particle Reynolds number $Re_p \ll 1$, the particle relaxation time is

$$\tau_{p,0} = \frac{\rho_p \, d_p^2}{18 \, \mu_f},$$ (2.12)

where $\rho_p$ is the particle density. For non-Stokesian particles, a correction factor $\psi(Re_p)$ is required, leading to

$$\tau_p = \tau_{p,0} \, \psi(Re_p) \,,$$ (2.13)

with

$$\psi(Re_p) = \frac{24}{Re_p} \int_0^{Re_p} \frac{dRe'}{C_D(Re')Re'}$$ (2.14)

| Method | Angle of attack | Induction factors |
|--------|-----------------|-------------------|
| *2D Ind* | $\alpha_{\mathrm{aero}}$ | From BEMT / 3D $C_P$ matching |
| *2D Geom* | $\alpha_{\mathrm{geom}}$ | Neglected |

Table 1. Summary of nomenclature and characteristics of the 2D solutions.

as first introduced by Israel & Rosner (1982). Thus, it is possible to express *Stk* as

$$Stk = Stk_0 \, \psi(Re_p), \tag{2.15}$$

where $Stk_0 = \tau_{p,0}/\tau_f$. We will first use $Stk_0$ to obtain a magnitude estimate for *Stk*; in § 4, this assumption will be removed.

The fluid characteristic time can be expressed differently according to the phenomenon of interest. Depending on the choice, a definition of the Stokes number follows. Given a time-averaged solution, e.g., the output of RANS equations, the local characteristic time scale of the flow can be defined as $\tau_{f,\mathrm{loc}} = 1/||\nabla v_f||_2$ (we consider the Frobenius norm). On an airfoil section, the time scale can be defined as $\tau_{f,\mathrm{sec}} = c/V_{\mathrm{rel}}$, where $c$ is the chord of the section and $V_{\mathrm{rel}}$ is the relative velocity between the section and the fluid. Also, it is possible to define a time scale $\tau_{f,\mathrm{ind}}$ related to the flow in the stream tube upstream of the rotor, where the induction is generated. Following these time scales, we can define a local, sectional, and induction Stokes number as

$$Stk_{\mathrm{loc}} = \tau_p ||\nabla v_f||_2 \,, \qquad Stk_{\mathrm{sec}} = \tau_p \frac{V_{\mathrm{rel}}}{c} \,, \text{ and } \qquad Stk_{\mathrm{ind}} = \frac{\tau_p}{\tau_{f,\mathrm{ind}}} \,, \tag{2.16}$$

respectively, which will be useful for the analysis of the results.

The region of the unperturbed flow where particles are released is characterised by a maximum local Stokes number $Stk_{\mathrm{loc},0} \sim 0.2$, at a minimum of three chords upstream of the sections in 2D or one radius upstream of the rotor in 3D. The constraint on $Stk_{\mathrm{loc}}$ implies that large particles are released further upstream since particles are initialised in local equilibrium with the flow field.

## 3. Limiting solutions

The carrying and dispersed phases are computed on two geometries, i.e., a wind turbine rotor and a small-scale propeller. The ratio of the rotor diameters is 33.5. The 3D solution is compared with the 2D limiting solutions introduced in figure 1c and figure 1d, whose characteristics are summarised in table 1. Details of 2D and 3D grid discretisation and convergence analysis are reported in Appendix B.

### 3.1. *Wind turbine rotor*

The rotor under analysis was studied in the Unsteady Aerodynamics Experiment (UAE) Phase VI (Hand *et al.* 2001*a*). This experiment has been a benchmark for understanding the three-dimensional aerodynamics of rotors and improving predictions of low-fidelity aerodynamic solvers (Jonkman 2003). The experiment studied a stall-controlled two-bladed wind turbine with a diameter of 10.058 m and was carried out in the NASA Ames Research Center wind tunnel. The test case chosen for this study belongs to Sequence S of the experimental campaign, where the rotor was upwind with no cone angle, the blade tip pitch was set to 3°, and the rotational speed was 72 rpm. Only the case with no yaw error is considered in the current work. We carry out simulations with a free-stream wind speed of $7\,\text{m s}^{-1}$. The tip-speed ratio TSR $= \Omega R / V_\infty$ is 5.4.

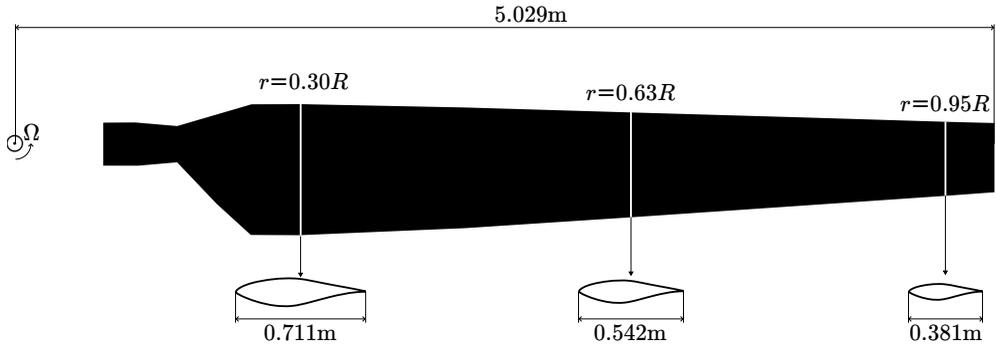

Figure 2. Visualisation of the blade plan-form and location of the reference 2D sections.

| $r/R$ | $\alpha_{\text{aero}}$ | $\alpha_{\text{geom}}$ |
|-------|------------------------|------------------------|
| 0.30  | 7.5°                   | 12.5°                  |
| 0.63  | 6.8°                   | 10.3°                  |
| 0.96  | 4.2°                   | 7.6°                   |

Table 2. Angle of attack of the wind turbine sections with ($\alpha_{\text{aero}}$, *2D Ind*) and without ($\alpha_{\text{geom}}$, *2D Geom*) induced velocities at $V_\infty = 7\,\text{m s}^{-1}$.

Denoting the radial distance from the rotation axis with $r$ and the rotor radius with $R$, the blade has a cylindrical root for $0.101 < r/R < 0.176$, a transition region for $0.176 < r/R < 0.250$, and a tapered and twisted aerodynamic region for $r/R > 0.250$ using an S809 airfoil. The comparison is carried out at three radial locations, namely $r/R = 0.30$, $r/R = 0.63$, and $r/R = 0.95$. These are shown in figure 2 on the blade plan-form. Experimental measurements of the pressure distributions were also available in these sections (Hand *et al.* 2001*b*) and are therefore included in the analysis of the carrying phase.

The induced velocity on the rotor sections is retrieved by solving BEMT in OpenFAST (2024) to compute $\alpha_{\text{aero}}$ in the *2D Ind* case. The numerical model used is publicly available in the code's GitHub repository. It includes a set of aerodynamic coefficients for nine different radial positions, obtained by extrapolating with the Viterna model and correcting the measurements on the S809 airfoil executed at the Ohio State University (Ramsay *et al.* 1995). Results are reported in table 2 for the three radial stations under analysis together with $\alpha_{\text{geom}}$ for the *2D Geom* case. These are used to compute the flow field around the sections to then simulate the transport of the dispersed phase. For completeness, in table 3 the thrust and torque computed with the RANS simulation are compared with the BEMT results and the experimental measurements.

The pressure coefficient of the 2D and 3D simulations are shown in figure 3 and compared with the experimental measurements. As expected, the *2D Geom* case leads to wrong results. Conversely, *2D Ind* leads to good agreement with both the 3D simulation and the experiments. The agreement with the 3D simulation is particularly good at the pressure side for $x/c < 0.5$, which is the most relevant part for the computation of particle impingement. On the suction side, the flow expands more and lower values of pressure are found, as recently observed also by Oztekin & Bain (2024). Overall, the 3D simulation is accurate both on the suction and on the pressure side and is generally in better agreement with the experiments when compared to 2D simulations.

| Method | Thrust (N) | Torque (N m) |
|---|---|---|
| RANS | 1150 | 766 |
| BEMT | 1224 | 779 |
| Exp. | $1150 \pm 17$ | $796 \pm 17$ |

Table 3. Comparison of thrust and torque computed with the different numerical models and the experimental measurements, expressed as mean value ± one standard deviation.

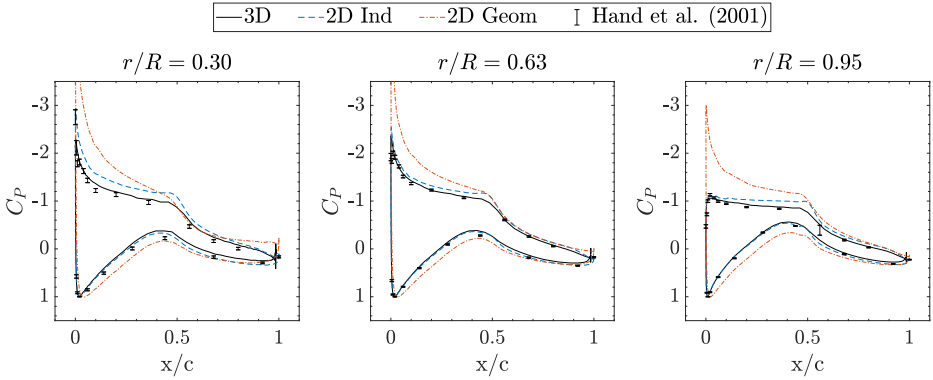

Figure 3. Pressure coefficient from the 3D and 2D flow-fields used to compute the discrete phase. The error-bar (⊺) represents one standard deviation from the mean value of the pressure tap experimental acquisition.

| $d_p$ (μm) | $\tau_{p,0}$ (s) | $Stk_{\mathrm{sec},0}\vert_{0.30R}$ | $Stk_{\mathrm{sec},0}\vert_{0.63R}$ | $Stk_{\mathrm{sec},0}\vert_{0.95R}$ |
|---|---|---|---|---|
| 20 | $1.2 \times 10^{-3}$ | $2.3 \times 10^{-2}$ | $5.6 \times 10^{-2}$ | $1.2 \times 10^{-1}$ |
| 63 | $1.2 \times 10^{-2}$ | $2.3 \times 10^{-1}$ | $5.6 \times 10^{-1}$ | $1.2 \times 10^{0}$ |
| 200 | $1.2 \times 10^{-1}$ | $2.3 \times 10^{0}$ | $5.6 \times 10^{0}$ | $1.2 \times 10^{1}$ |
| 630 | $1.2 \times 10^{0}$ | $2.3 \times 10^{1}$ | $5.6 \times 10^{1}$ | $1.2 \times 10^{2}$ |

Table 4. Sectional Stokes number $Stk_{\mathrm{sec},0}$ of the simulations at $7\,\mathrm{m\,s^{-1}}$ for four droplet diameters at the three radial stations $r/R = 0.30$, $r/R = 0.63$, and $r/R = 0.95$. The time scale of the carrying phase at the sections is $\tau_{f,\mathrm{sec}}\vert_{0.30R} \sim 0.05\,\mathrm{s}$, $\tau_{f,\mathrm{sec}}\vert_{0.63R} \sim 0.02\,\mathrm{s}$, $\tau_{f,\mathrm{sec}}\vert_{0.95R} \sim 0.01\,\mathrm{s}$.

### 3.1.1. *Discrete phase*

The discrete phase is computed for different particle diameters. The values presented in this section are listed in table 4 with the sectional Stokes number $Stk_{\mathrm{sec},0}$ computed with equation (2.16) considering $\tau_p = \tau_{p,0}$.

The section-normalised collection efficiency is computed with equation (2.10) and reported in figure 4. The horizontal axis is the normalised curvilinear abscissa of the section, whose value is 0 at the leading edge and ±1 at the trailing edge, with negative values on the pressure side and positive values on the suction side. The figure shows the radial variation of the collection efficiency for a fixed droplet diameter on each row, and the variation due to droplet size at a fixed radial station on each column. Moreover, the sectional Stokes number $Stk_{\mathrm{sec},0}$ increases monotonically as reported in table 4, although the sections have different angles of attack.

Considering the smallest particles ($d_p = 20\,\mu\mathrm{m}$), at $r/R = 0.30$, both the 3D and the *2D Ind* solutions present the minimum values of collection efficiency, whereas the higher angle of attack of *2D Geom* leads to larger velocity gradients, which prevent particles from hitting the section. At larger radial stations, 3D results show the best agreement with the *2D Ind* simulations. The same occurs considering $d_p = 63\,\mu\mathrm{m}$ (second row), where the difference between *2D Ind* and *2D Geom* increases. The agreement or mismatch between 2D and 3D can be seen as a direct consequence of the accuracy of the $C_P$ distribution shown in figure 3, which is representative of the external velocity of the section.

Larger particles ($d_p = 200\,\mu\mathrm{m}$) still show a good agreement between the *2D Ind* simulation and the 3D solution for $r/R = 0.30$ and $r/R = 0.63$. It is worth noticing that, at $r/R \geq 0.63$, both the 2D simulations have almost reached their ballistic limit (equation 2.11) on the pressure side, represented in the plots with a dotted line (··). This means that the particles are almost undeflected by the flow field around the section; such a phenomenon is expected for $Stk_{\mathrm{sec}} \gg 1$. For $r/R = 0.95$, the 3D solution is located between the 2D ones, i.e., it is beyond the *2D Ind* ballistic limit (computed with $\alpha_{\mathrm{aero}}$) and starts converging to the *2D Geom* one (computed with $\alpha_{\mathrm{geom}}$). The behaviour started with smaller droplets, and continues until the 3D solution reaches the *2D Geom* ballistic solution at $d_p = 630\,\mu\mathrm{m}$. The same trend is found at the two smaller radial locations; the *2D Geom* ballistic limit will be reached last at the smallest radial location.

We have presented the results considering the whole blade for each particle diameter; however, the trends are clearer when looking at a single section for increasing particle size. When particles are sufficiently small, the *2D Ind* approach provides a good approximation of the 3D solution. As the particle diameter increases and $Stk_{\mathrm{sec}} \gg 1$, the solution becomes ballistic; this is clearly seen in both *2D Ind* and *2D Geom* cases. The 3D solution departs from the *2D Ind* solution and converges to the *2D Geom* one for increasing particle size. This, however, occurs earlier at the tip of the blade; the smaller is the radial coordinate, the larger is the particle diameter required to reach the *2D Geom* ballistic solution in 3D.

We also present the relative particle-section velocity vector $\boldsymbol{v}_p - \boldsymbol{\Omega} \times \boldsymbol{r}$ at impact in figure 5 for $d_p = 63\,\mu\mathrm{m}$ and $d_p = 630\,\mu\mathrm{m}$. This parameter can be relevant in ice accretion and erosion models. In the first two rows, we show the magnitude of the vector normalised with the ballistic velocity. The magnitude from 2D simulations is generally better captured using the *2D Ind* approach; however, the impingement limits of the droplets on the section of the 3D simulation become more similar to the *2D Geom* case when the particles are sufficiently large. In the last two rows, the angle of the in-plane component of the vector is computed with respect to the section chord. In this case, the trend is more similar to the collection efficiency. In 2D, the impact angle approaches the value given by the aerodynamic or geometric angle of attack of the section as the particle size increases. When droplets are sufficiently large, the 3D solution approaches the *2D Geom* solution and impact the section with an angle $\alpha_{\mathrm{geom}}$ with respect to the chord.

We will further discuss the results after presenting those on the propeller.

### 3.2. *Small-scale propeller*

The small-scale propeller considered is a three-bladed rotor using VarioProp 12C blades. The diameter of the rotor is 300 mm and the pitch angle measured at $0.75R$ is $26.5°$. It features a spinner with a diameter of 65 mm and a nacelle of 270 mm length and is shown in figure 6. The propeller was tested in the De Ponte wind tunnel of Politecnico di Milano by Zanotti & Algarotti (2022) and the blade geometry was provided from a 3D scan (Piccinini *et al.* 2020). A fixed rotational speed of 7050 rpm was used to obtain a tip Mach number of 0.325 to reproduce that of a full-scale electric vertical take-off and landing (eVTOL) aircraft in cruise conditions. Two propeller advance ratios were considered, i.e., $J = 0.4$ and $J = 0.8$; the advance ratio $J$ is defined as $J = V_\infty/(nD)$, where $n$ is the rotational speed measured in Hz. The two flight conditions are obtained with $V_\infty = 14.1\,\text{m s}^{-1}$ and $V_\infty = 28.2\,\text{m s}^{-1}$, respectively, and the corresponding tip-speed ratios are TSR = 7.9 and TSR = 3.9. Freestream temperature and density are 294.05 K ad 1.18 kg m$^{-3}$. Experimental measurements of thrust and torque are used for comparison with RANS simulations as shown in table 5. We presented a preliminary analysis at $J = 0.8$ in Caccia & Guardone (2025).

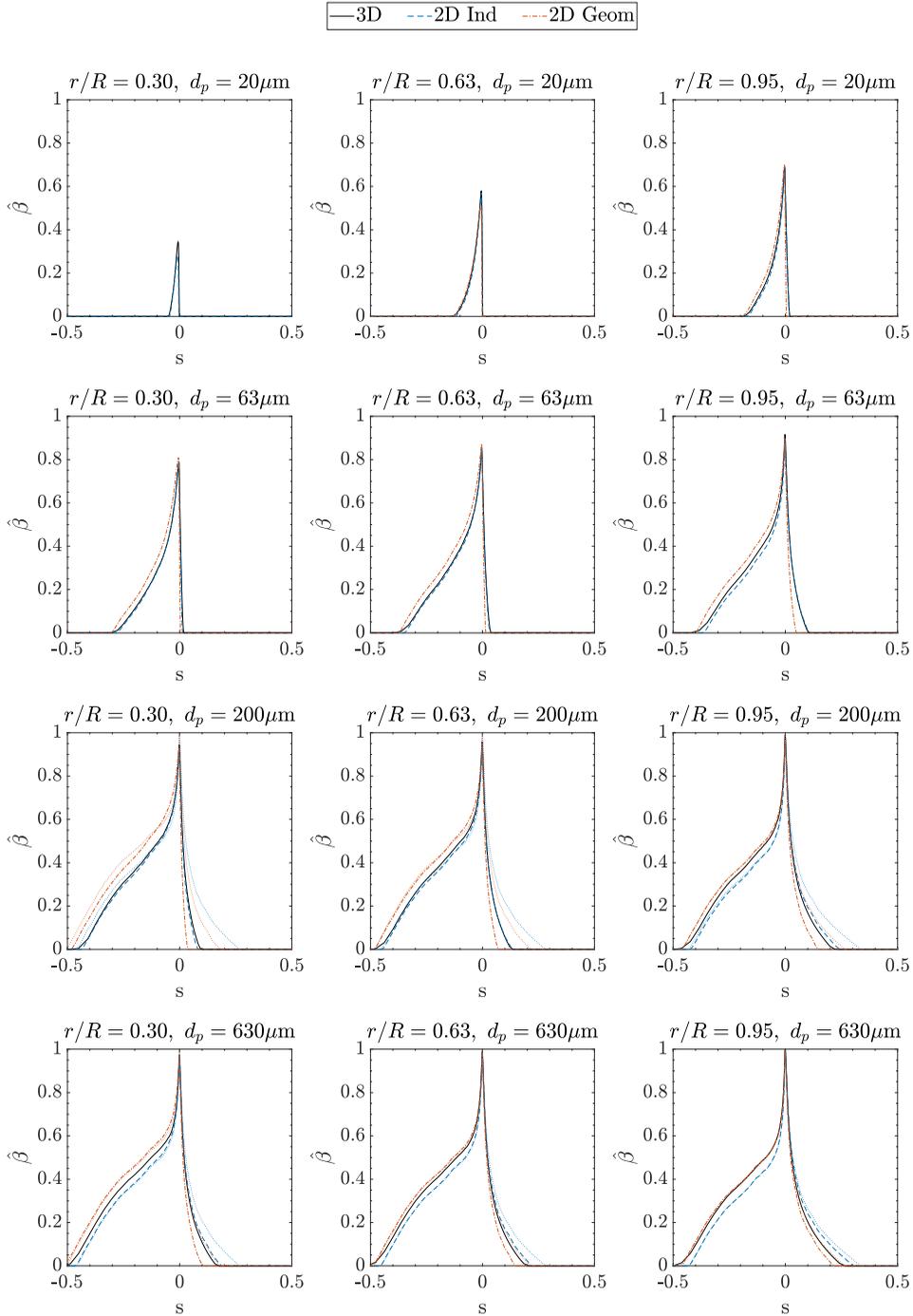

Figure 4. Section-normalised collection efficiency $\hat{\beta}$ as a function of the normalised curvilinear abscissa $s$ of the section. The curvilinear abscissa is $s = 0$ on the leading edge, $s = \pm 1$ on the trailing edge, with negative values on the pressure side and positive values on the suction side. Each column represents a radial location ($r/R = 0.30$, $r/R = 0.63$, $r/R = 0.95$). Each row represents a droplet size ($d_p = 20\,\mu\mathrm{m}$, $d_p = 63\,\mu\mathrm{m}$, $d_p = 200\,\mu\mathrm{m}$, $d_p = 630\,\mu\mathrm{m}$). In the last two rows, the dotted lines ($\cdot\cdot$) represent the ballistic limit of the solution considering $\alpha_{\mathrm{aero}}$ (*2D Ind*) and $\alpha_{\mathrm{geom}}$ (*2D Geom*).

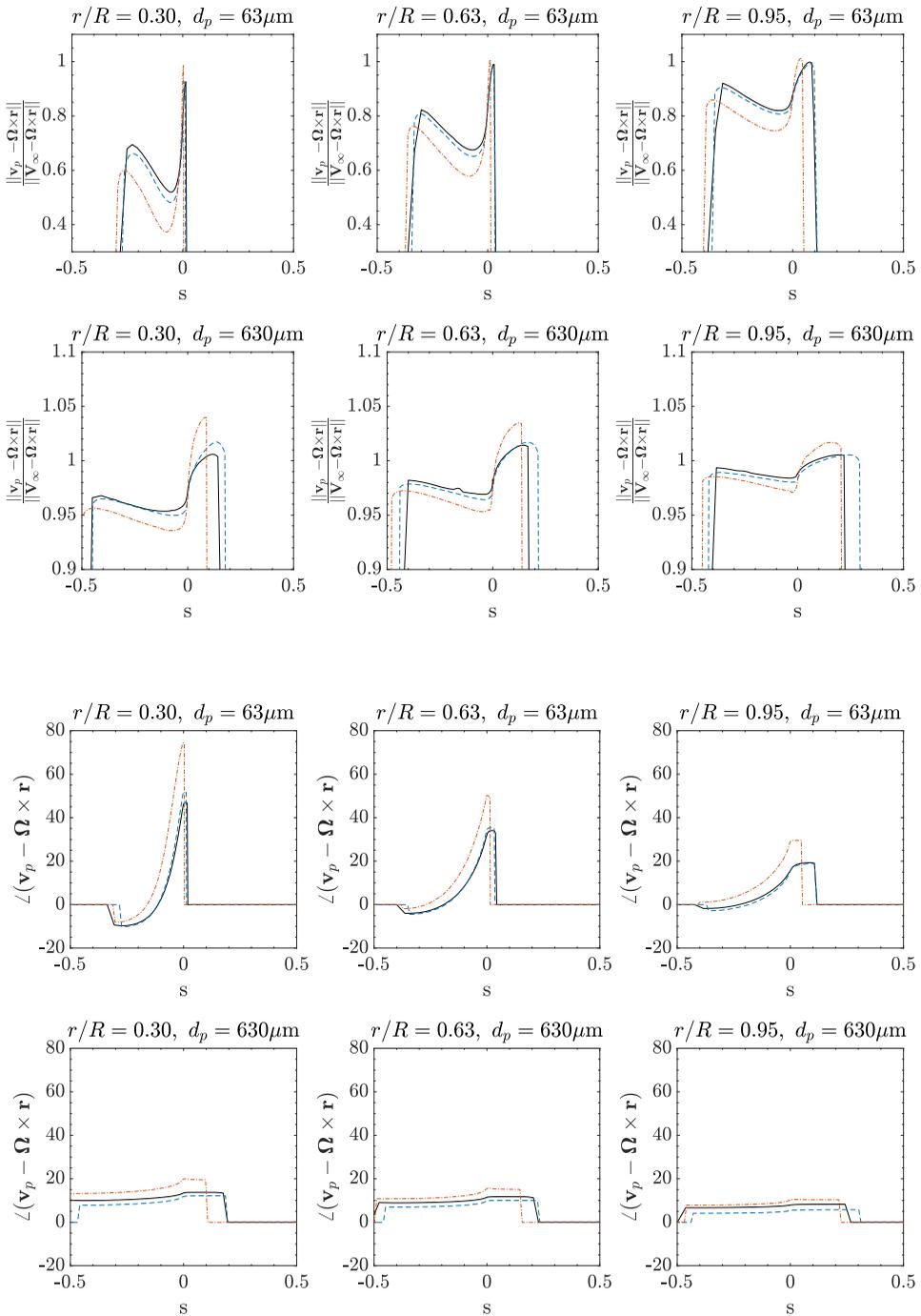

Figure 5. Particle-surface relative velocity at impact as a function of the normalised curvilinear abscissa $s$ of the section. Each column represents a radial location ($r/R = 0.30$, $r/R = 0.63$, $r/R = 0.95$). First two rows: non-dimensionalised impact velocity ($d_p = 63\,\mu m$, $d_p = 630\,\mu m$). Last two rows: angle of the in-plane component with respect to the section chord ($d_p = 63\,\mu m$, $d_p = 630\,\mu m$).

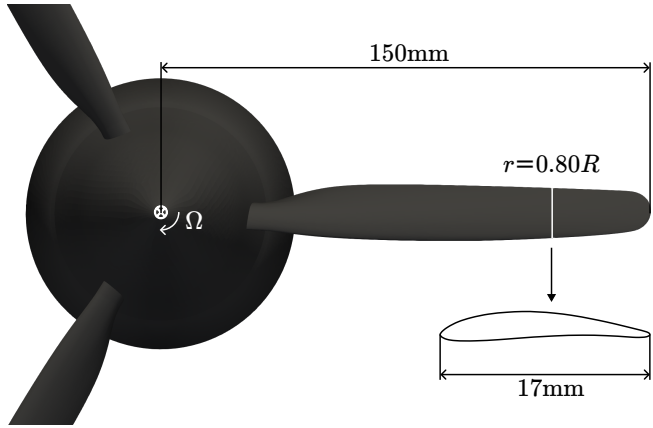

Figure 6. Front view of the propeller and section geometry.

| Method | Thrust$_{J=0.4}$ (N) | Thrust$_{J=0.8}$ (N) | Torque$_{J=0.4}$ (N m) | Torque$_{J=0.8}$ (N m) |
|---|---|---|---|---|
| RANS | 18.7 | 13.8 | 0.84 | 0.73 |
| Exp. | 20.9 | 13.6 | 0.83 | 0.71 |

Table 5. Computed torque and thrust of the propeller and comparison with the experimental data. Data include the blades and the spinner, but not the hub.

| $J$ | $\alpha_{\text{aero}}$ | $\alpha_{\text{geom}}$ |
|---|---|---|
| 0.4 | 10.0° | 17.1° |
| 0.8 | 4.5° | 8.5° |

Table 6. Angle of attack of the propeller section ($r/R = 0.8$) with ($\alpha_{\text{aero}}$) and without ($\alpha_{\text{geom}}$) induced velocity at the two advance ratios $J$ under analysis.

The comparison is carried out at the radial location $r/R = 0.8$ as shown in figure 6. The section chord is $c = 17\,\text{mm}$. The sectional aerodynamic angle of attack $\alpha_{\text{aero}}$ for the *2D Ind* case is chosen to match the pressure coefficient of the 3D solution on the pressure side (Narducci & Kreeger 2012; Narducci *et al.* 2012; Oztekin & Bain 2024). Indeed, from the previous results, this proved a necessary condition to obtain an accurate collection efficiency at small values of sectional Stokes number. The value is shown in table 6 together with the geometric angle of attack of the *2D Geom* case.

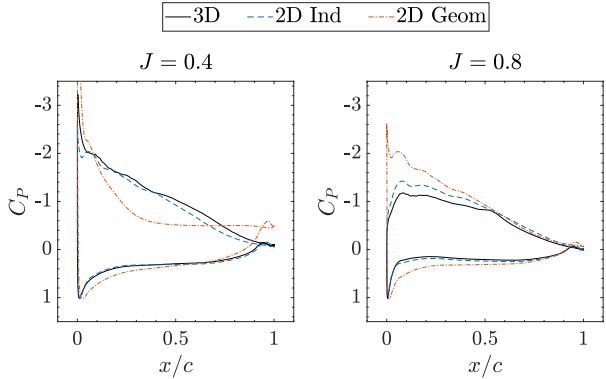

Figure 7. Pressure coefficient from the 3D and 2D flow-fields used to compute the discrete phase.

| $d_p$ (μm) | $\tau_{p,0}$ (s) | $Stk_{sec,0}$ |
|---:|---:|---:|
| 2.5 | $2.0 \times 10^{-5}$ | $1 \times 10^{-1}$ |
| 8.0 | $2.0 \times 10^{-4}$ | $1 \times 10^{0}$ |
| 25.0 | $2.0 \times 10^{-3}$ | $1 \times 10^{1}$ |
| 80.0 | $2.0 \times 10^{-2}$ | $1 \times 10^{2}$ |

Table 7. Average sectional Stokes number $Stk_{sec,0}$ of the simulations at $J = 0.4$ and $J = 0.8$ for four droplet diameters at the radial station $r/R = 0.80$. The time scale of the carrying phase at the section is $\tau_{f,sec}|_{J=0.4} = 1.9 \times 10^{-4}$s and $\tau_{f,sec}|_{J=0.8} = 1.8 \times 10^{-4}$s. An average value was used to compute $Stk_{sec,0}$.

The pressure coefficient of the 3D and 2D simulations is compared in figure 7. With the chosen $\alpha_{aero}$, a good match on the whole pressure side of the section is obtained for both advance ratios. It is worth noticing that, due to the low Reynolds number, fully-turbulent simulations were required in 2D to match the pressure coefficient. Indeed, the laminar separation bubble in the 2D simulations leads to large separation zones on both the blade's pressure and suction sides, not representative of the behaviour on a rotating component.

### 3.2.1. *Discrete phase*

The discrete phase is computed for different particle diameters. The values presented in this section are listed in table 7 with the an average sectional Stokes number $Stk_{sec,0}$ between the two advance ratios, computed with equation (2.16) considering $\tau_p = \tau_{p,0}$.

The section-normalised collection efficiency is computed with equation (2.10) and reported in figure 8. The figure shows the collection efficiency for a fixed droplet diameter, changing the section angle of attack (advance ratio), on each row, and the variation due to droplet size at a fixed angle of attack on each column. Since $\alpha_{aero}(J = 0.4) \sim \alpha_{geom}(J = 0.8)$, the corresponding curves in the two columns are similar.

The same trends observed on the wind turbine rotor are retrieved and thus can be summarised as follows. When $Stk_{sec}$ is sufficiently small, the 3D solution can be well approximated with the *2D Ind* approach. With increasing droplet size and Stokes number, the 2D solutions reach their respective ballistic limit, computed with equation 2.11 and represented in the plots with a dotted line ($\cdots$) in the last two rows; the 3D solution transitions from the *2D Ind* to the *2D Geom* solution, ultimately reaching the ballistic limit of the *2D Geom* solution. It is worth noticing that the collection efficiency reaches the ballistic solution at a different rate on the suction and pressure sides; moreover, the local collection efficiency can exceed that of the ballistic solution.

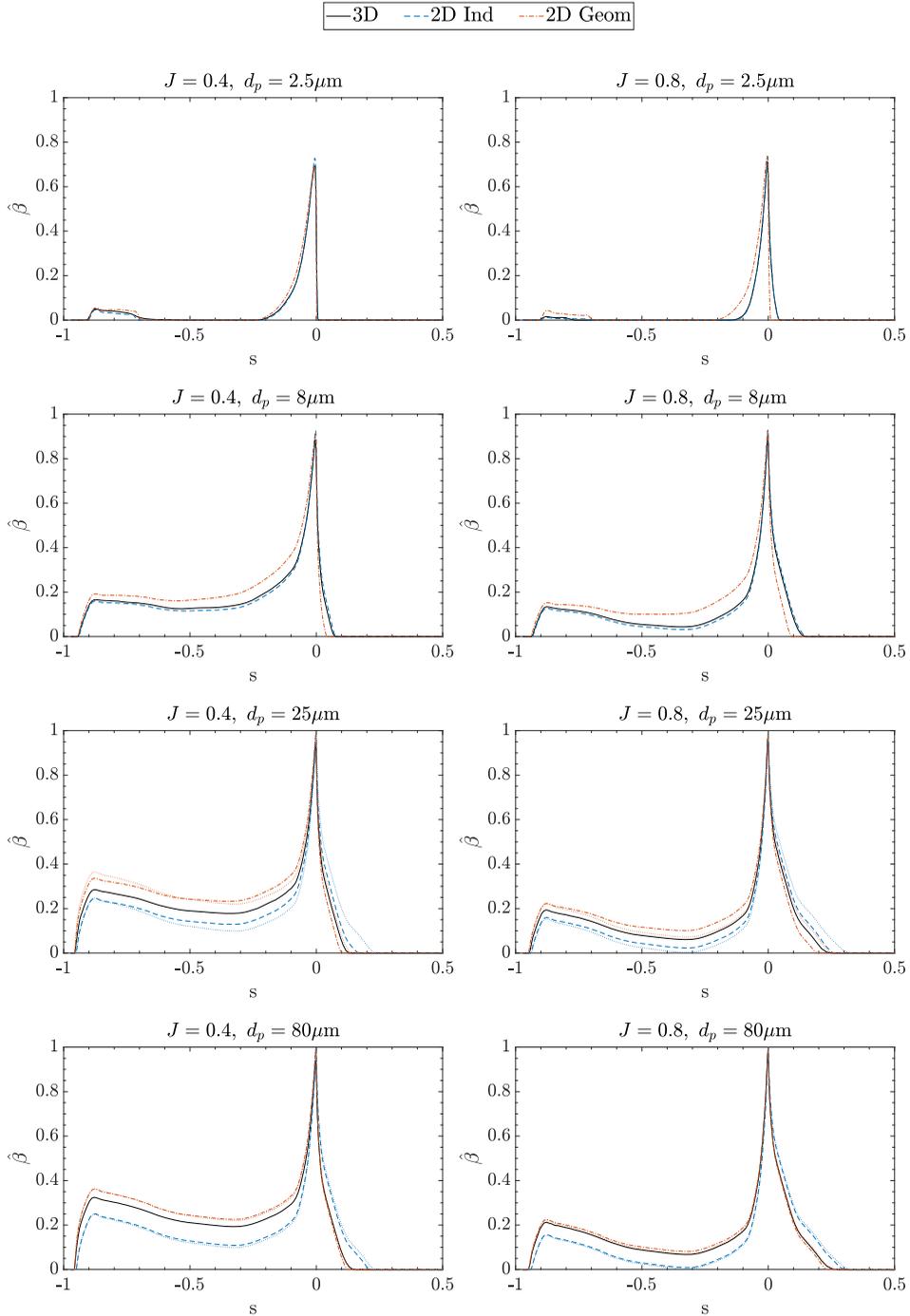

Figure 8. Section-normalised collection efficiency $\hat{\beta}$ as a function of the normalised curvilinear abscissa $s$ of the propeller section at $r/R = 0.80$. The curvilinear abscissa is $s = 0$ on the leading edge, $s = \pm 1$ on the trailing edge, with negative values on the pressure side and positive values on the suction side. Each column represents a propeller advance ratio ($J = 0.4$, $J = 0.8$). Each row represents a droplet size ($d_p = 2.5\,\mu$m, $d_p = 8\,\mu$m, $d_p = 25\,\mu$m, $d_p = 80\,\mu$m). In the last two rows, the dotted lines ($\cdot\cdot$) represent the ballistic limit of the solution considering $\alpha_{\text{aero}}$ (*2D Ind*) and $\alpha_{\text{geom}}$ (*2D Geom*).

The relative particle-section velocity vector $\boldsymbol{v}_p - \boldsymbol{\Omega} \times \boldsymbol{r}$ at impact are presented in figure 9 for for $d_p = 8\,\mu\text{m}$ and $d_p = 80\,\mu\text{m}$ in terms of magnitude (first two rows) and angle with respect to the section chord (last two rows). The trends are similar to those observed on the wind turbine, also for these variables. In 2D simulations, the magnitude is generally better captured when the droplet is immersed in the *2D Ind* flow field, although the impingement limits for large particles are not captured correctly. The agreement for the impact direction follows the trends of the collection efficiency.

It is worth looking at the dip in velocity magnitude at $J = 0.8$. In the *2D Ind* case, at $J = 0.8$, the collection efficiency was $\hat{\beta} \sim 0$ on the pressure side at $s \sim -0.3$, non-null for $d_p \gtrsim 8\,\mu\text{m}$. Part of the particles impinging on the section at $s < -0.3$ pass through the boundary layer of the section, causing a drop in the particle speed. A good agreement between the 2D simulation and the 3D solution is found for $d_p = 8\,\mu\text{m}$; however, for $d_p = 80\,\mu\text{m}$, the droplet velocity is affected by the passage in the boundary layer only in the *2D Ind* case, and is unaffected both in the 3D solution and in the *2D Geom* simulation. The passage of particles through the boundary layer does not affect their impact angle.

### 3.3. *Assessment of limiting solutions*

The results presented until now highlighted some relations between three-dimensional multi-phase rotor flow fields and their 2D approximation. First, for small-enough particles, the 3D behaviour can be represented through a 2D solution if the pressure distribution on the part of the section exposed to the droplets is accurate. Indeed, this is representative of the pressure outside the boundary layer and thus of the external velocity field, which carries the particles. A necessary condition for this to occur is that the sectional angle of attack in 2D and 3D is similar. In general, this is not sufficient. The different behaviour of 2D and 3D boundary layers on a rotating component required a fully turbulent simulation of the 2D propeller section to obtain a compliant $C_P$. Second, for large sectional Stokes number ($Stk_{\text{sec},0} \gg 1$), two limiting ballistic solutions exist in the 2D simulations, given by the different orientations of the sections with respect to the free-stream wind. The ballistic regime reached by particles in the 3D simulations is the geometric one, not accounting for the induced velocity vector. Third, the ballistic solution may be reached at a different rate on the suction and pressure sides, and local convergence to the ballistic solution can be non-monotonic. The convergence rate of 3D and 2D simulations to their respective ballistic solution is apparently different.

We analyse the error of the collection efficiency of the 2D simulations with and without induction with respect to the 3D solution. The error is defined as

$$\text{err}(\hat{\beta}) = \frac{\int |\hat{\beta}_{2\text{D}} - \hat{\beta}_{3\text{D}}| ds}{\int \hat{\beta}_{3\text{D}} ds} \ , \tag{3.1}$$

where $\hat{\beta}_{2\text{D}}$ and $\hat{\beta}_{3\text{D}}$ are the 2D and 3D collection efficiency at a specific radial coordinate. The results are reported in figure 10 as a function of the sectional Stokes number.

Let us follow the *2D Ind* case first. For the smallest particles, resulting in $Stk_{\text{sec},0} < 10^{-1}$, the normalised error on the collection efficiency can be large. The dimensional error, however, can be small since the impinging mass on the blade is either small or even null for such a small Stokes number. Second-order effects, such as local spanwise flows, might become relevant and lead to differences even in 2D simulations that correctly capture the overall external flow ahead of the section. For $Stk_{\text{sec},0} > 10^{-1}$, a plateau of limited error might exist. Particles filter out such second-order effects, and the 2D approximation leads to good results. Afterwards, the error builds up until it reaches a constant value when both the 2D and 3D solutions reach their respective ballistic limit. Conversely, when *2D Geom* is used, there might be an initial plateau of large error. Afterwards, the error decreases until it reaches a minimum, i.e., when the 3D and 2D solutions reach their ballistic limit.

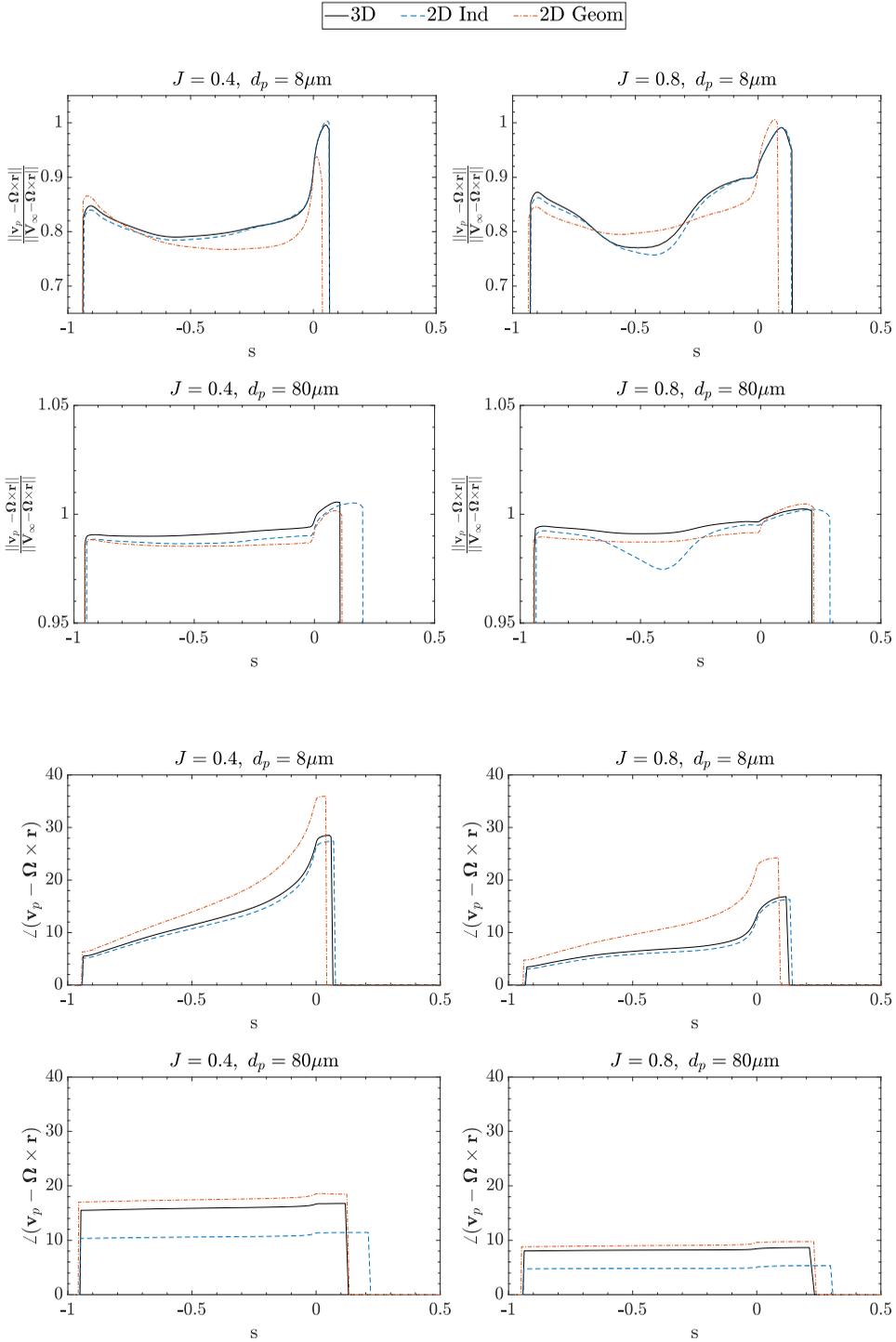

Figure 9. Particle-surface relative velocity at impact as a function of the normalised curvilinear abscissa $s$ of the section. Each column represents a propeller advance ratio ($J = 0.4$, $J = 0.8$). First two rows: non-dimensionalised impact velocity ($d_p = 8\,\mu\text{m}$, $d_p = 80\,\mu\text{m}$). Last two rows: angle of the in-plane velocity component with respect to the section chord ($d_p = 8\,\mu\text{m}$, $d_p = 80\,\mu\text{m}$).

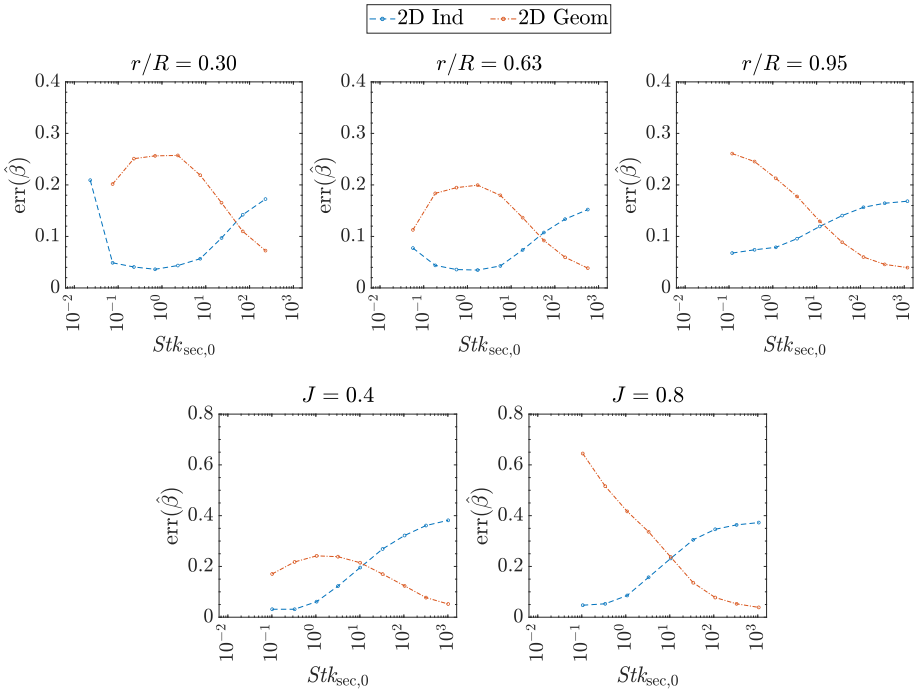

Figure 10. Error of 2D simulations with respect to the 3D solution, computed with equation (3.1), as a function of the sectional Stokes number. Top: three wind turbine sections ($r/R = 0.30$, $r/R = 0.63$, $r/R = 0.95$) at fixed operating conditions. Bottom: propeller section at $r/R = 0.80$ at different advance ratios ($J = 0.4$, $J = 0.8$).

Between the two 2D approaches, there is a clear transition region, where the full solution cannot be represented accurately by the 2D limiting solutions; indeed, for increasing particle diameter, the error increases when the aerodynamic angle of attack is used, whereas the error associated with the geometric angle of attack decreases. In this transition region, the particles in the 3D case change behaviour: released far upstream of the rotor in an unperturbed region, a particle larger than those in equilibrium with the induced velocity and smaller than the ballistic ones is only partially affected by the induced velocity field. However, relating this behaviour to the sectional Stokes number would be misleading, as the fluid time scale at which this phenomenon occurs does not have a direct relation with the sectional one $\tau_{f,\text{sec}} = c/V_{\text{rel}}$. Indeed, the transition region is located at different orders of $Stk_{\text{sec},0}$ in the cases analysed, and ranges roughly between $10^0$ and $10^3$. However, one would expect it to start at $Stk \sim 0.1$, be centred at $Stk \sim 1$, and end at $Stk \sim 10$.

We provide an estimate of a more meaningful fluid time scale $\tau_f$, by replacing the length scale in $\tau_{f,\text{sec}}$ with the rotor radius $R$, as induction is built upstream of the rotor and not locally on the section. Thus, we can estimate an induction Stokes number as $Stk_{\text{ind},0} \sim \tau_{p,0}V_{\text{rel}}/R$. We then present in figure 11 the normalised velocity $v_p/V_\infty$ and Reynolds number $Re_p$ of particles of different diameters as a function of time to visualise their behaviour upstream of the rotor for different $Stk_{\text{ind},0}$. Results are presented for the wind turbine section at $r/R = 0.95$ and the propeller section at $J = 0.8$. Time is non-dimensionalised with the rotor's rotation period, and $t = 0$ corresponds to the particles' impact on the section. Vertical dashed lines represent a blade passage in front of the particles. From the plot, we naturally see that the amount of induction created decays with the distance from the rotor disk; it is also clear that the induced velocity increases at every blade passage. The phenomenon becomes less impulsive at smaller radial stations due to the increasing influence of the other blades (the figure is not reported for brevity). Large droplets are almost unaffected by the induced flow field, and the transition region identified earlier is found at $0.1 \lesssim Stk_{ind,0} \lesssim 100$. It is noticeable that, for $Stk_{ind,0} \gg 1$, also $Re_p \gg 1$. The effective $Stk_{ind}$, accounting for $\psi(Re_p) < 1$ of equation (2.13) would be smaller: e.g., $\psi(10^3) \sim 0.1$, and thus the transition region would be located in $0.1 \lesssim Stk_{ind} \lesssim 10$. We note, however, that the values reported in figure 11 at $t \sim 0$ also include the effect of the section and its local flow field, and not only the effect of induction.

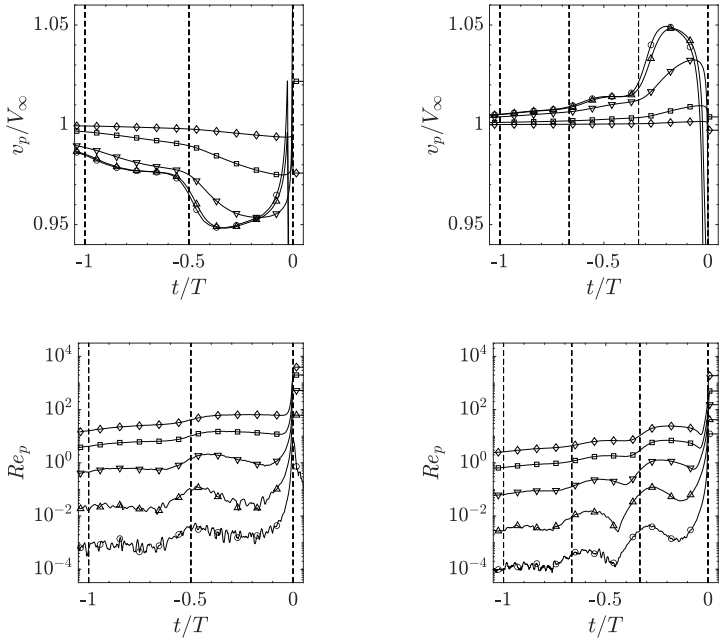

Figure 11. Normalised velocity magnitude (top) and Reynolds number (bottom) of a particle in the 3D flow field upstream of the rotor. The rotor is impacted at $t = 0$. The vertical dashed lines ($--$) represent a blade passage. The Stokes number $Stk_{ind}$ is computed assuming $\psi(Re_p) = 1$ Left: wind turbine ($r_p(t = 0) = 0.95R$). Right: propeller ($r_p(t = 0) = 0.80R$).

## 4. A model of the delayed response of a particle to the rotor-induced velocity

We have shown that the non-equilibrium state of the particles upstream of the rotor disk leads to an intermediate regime, where particles reach the section following neither its aerodynamic nor its geometric angle of attack. Here, we present a simple analytical model of the particles' response to the induced velocity field in the stream tube upstream of the rotor disk. The model relies exclusively on the knowledge of the rotor operating conditions and the aerodynamic and geometric angles of attack of the section, i.e., on the axial and tangential induction factors.

Let us consider a particle in initial equilibrium with the free-stream velocity immersed in a carrying velocity field $\mathbf{v}_f(x, r) = \mathbf{V}_\infty + \mathbf{v}_{f,\mathrm{ind}}(x, r)$ where $\mathbf{v}_{f,\mathrm{ind}}(x, r)$ is the velocity induced by the rotor, $x$ is the axial coordinate aligned with $\mathbf{V}_\infty$ and the rotation axis, and $r$ is the radial one. At $x = 0$, the particle impacts the rotor blade section at radial coordinate $r$. Here, the magnitude of the carrying phase induced velocity $||\mathbf{v}_{f,\mathrm{ind}}(0, r)||_2 = V_{f,\mathrm{ind}}$ is

$$V_{f,\mathrm{ind}} = \sqrt{(aV_\infty)^2 + (a'\Omega r)^2} \,, \tag{4.1}$$

where the axial and tangential induction factors $a$ and $a'$ are evaluated at the radial location $r$. The convention chosen is such that the induction factors are positive when they have the same direction as the axial and tangential velocity vectors, respectively, so that a rotor extracting momentum from the fluid will have $\{a < 0, \ a' > 0\}$ while a rotor adding momentum to the fluid will have $\{a > 0, \ a' < 0\}$. We look for a form of $\mathbf{v}_{f,\mathrm{ind}}(x, r)$ allowing a simple solution of the problem. The analytical solutions for an actuator disk in axis-symmetric conditions, as those derived by Conway (1995), might prove useful; for reference, the axial velocity component along the symmetry axis (with direction $\hat{\mathbf{x}}$) of a uniformly loaded actuator disk reads

$$\frac{\mathbf{v}_f(x, 0) \cdot \hat{\mathbf{x}}}{V_{f,\mathrm{ind}} \cdot \hat{\mathbf{x}}} = 1 + \frac{x/R}{\sqrt{1 + (x/R)^2}} \,, \tag{4.2}$$

where the characteristic length scale of the problem, $R$, is clearly visible. Such solutions approximate well the flow field upstream of a rotor for $x \lesssim -R$, where the influence of individual blades becomes negligible; this was shown on wind turbine rotors numerically (Medici *et al.* 2011; Troldborg & Meyer Forsting 2017), experimentally (Medici *et al.* 2011), and in field testing (Simley *et al.* 2016; Borraccino *et al.* 2017). Conversely, a particle experiences the highest gradients of velocity induction during discrete blade passages for $x > -R$. Thus, we propose a different description.

We consider a Lagrangian sampling of the carrying velocity field following a particle of the carried phase that will pass through the blade at a radial location $r$ and $t_p = 0$, where $t_p$ is the time of the particle that is sampling the velocity field. We approximate the induction buildup of the carrying phase as

$$v_{f,\mathrm{ind}}(t_p) = V_{f,\mathrm{ind}} \, e^{t_p / \tau_{f,\mathrm{ind}}} \,, \tag{4.3}$$

where $\tau_{f,\mathrm{ind}}$ represents the induction time scale of the carrying phase for the radial location under analysis. Please note that $v_{f,\mathrm{ind}}(t_p)$ is a scalar function representing the magnitude of the induced velocity vector, measured in the time domain of the particle $t_p \leq 0$. This function has two relevant features: it can provide a fair approximation of the actuator disk behaviour described in equation (4.2) and it allows a straightforward solution to the problem.

The particle is in equilibrium with the fluid at $t_p = -\infty$ where $v_{f,\mathrm{ind}}(-\infty) = 0$. The only force acting on the particle is drag due to the difference in velocity between the fluid and the particle, $v_{f,\mathrm{ind}}(t_p) - v_{p,\mathrm{ind}}(t_p)$. Thus, the induced velocity of the particle $v_{p,\mathrm{ind}}(t_p)$ evolves according to

$$\tau_{p,\mathrm{ind}} \frac{dv_{p,\mathrm{ind}}}{dt_p} + v_{p,\mathrm{ind}}(t_p) = v_{f,\mathrm{ind}}(t_p) \,, \tag{4.4}$$

where $\tau_{p,\mathrm{ind}}$ represents the relaxation time of the particle. Given the induced velocity $v_{f,\mathrm{ind}}(t_p)$ of equation (4.3), the solution is of the form $v_{p,\mathrm{ind}}(t_p) = AV_{f,\mathrm{ind}}e^{t_p / \tau_{f,\mathrm{ind}}}$. In

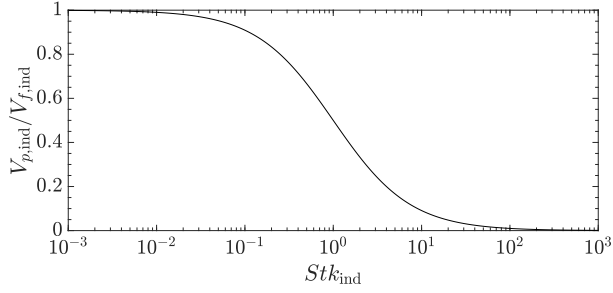

Figure 12. Response of a particle immersed in the induced velocity field $v_{f,\mathrm{ind}}(t_p) = V_{f,\mathrm{ind}}\, e^{t_p/\tau_{f,\mathrm{ind}}}$. The response is evaluated at $t_p = 0$, i.e., at the impact with the rotor section, as a function of the induction Stokes number $Stk_{\mathrm{ind}} = \frac{\tau_{p,\mathrm{ind}}}{\tau_{f,\mathrm{ind}}}$.

particular, the solution at impact $v_{p,\mathrm{ind}}(t_p = 0) = V_{p,\mathrm{ind}}$ is

$$V_{p,\mathrm{ind}} = \frac{V_{f,\mathrm{ind}}}{1 + Stk_{\mathrm{ind}}} \,, \tag{4.5}$$

where we have defined the induction Stokes number $Stk_{\mathrm{ind}} = \frac{\tau_{p,\mathrm{ind}}}{\tau_{f,\mathrm{ind}}}$. The response at $t_p = 0$ is a function of $Stk_{\mathrm{ind}}$ only, and is shown in figure 12. Once the magnitude $V_{p,\mathrm{ind}}$ is known, its direction will be the same as the induced velocity vector $\{aV_\infty,\ a'\Omega r\}$. Thus, the vector sum of the particle's induced velocity and the section's geometric velocity provides the relative velocity vector between the particle and the section. Both time scales defining $Stk_{\mathrm{ind}}$ still need to be determined.

The particle time scale, $\tau_{p,\mathrm{ind}}$, is given by the product of its time scale in the Stokesian regime $\tau_{p,0}$ (equation (2.12)) with a correction factor $\psi(Re_p)$ (equation (2.13)) to account for drag in the non-Stokesian regime, as suggested by Israel & Rosner (1982). Considering the Reynolds number of the particle at impact ($t_p = 0$)

$$Re_{\mathrm{ind}} = \frac{\rho_f\,\left|V_{f,\mathrm{ind}} - V_{p,\mathrm{ind}}\right| d_p}{\mu_f} \,, \tag{4.6}$$

the Stokes number at impact can be computed as

$$Stk_{\mathrm{ind}} = \frac{\tau_{p,\mathrm{ind}}}{\tau_{f,\mathrm{ind}}} = \frac{\tau_{p,0}\,\psi(Re_{\mathrm{ind}})}{\tau_{f,\mathrm{ind}}} \,. \tag{4.7}$$

A closed-form solution of $\psi(Re_p)$ was provided by Wessel & Righi (1988), using the drag coefficient relation suggested by Serafini (1954), valid for a spherical particle with $Re_p < 10^3$. After defining $k = \sqrt{0.158}$, the solution reads

$$\psi(Re_p) = 3\,\frac{k\,Re_p^{1/3} - \tan^{-1}\left(k\,Re_p^{1/3}\right)}{k^3\,Re_p} \,. \tag{4.8}$$

The last variable to determine is $\tau_{f,\mathrm{ind}}$, i.e., the induction time scale of the carrying phase, which must be modelled and defines the variation of induced velocity at the rotor plane, i.e., the maximum gradient of induced velocity. We express the time scale as the ratio between a length scale $L_{\mathrm{ref}}$ and a velocity scale $V_{\mathrm{ref}}$:

$$\tau_{f,\mathrm{ind}} = \frac{L_{\mathrm{ref}}}{V_{\mathrm{ref}}} \,. \tag{4.9}$$

We pick $L_{\text{ref}} = R$ as the length scale. For the velocity scale, we consider the relative velocity between the fluid and the section:

$$V_{\text{ref}} \sim \sqrt{(V_\infty \, (1+a))^2 + (\Omega r \, (1+a'))^2} \, . \tag{4.10}$$

We note that the particle velocity would be required, since $v_{f,\text{ind}}(t_p)$ is a function of $t_p$, i.e., it is sampled through the particle advection velocity; however, being this a scale, the difference is negligible. Similarly, one might use the freestream values; in this case, differences might arise for $V_\infty \sim 0$. Thus, we express the fluid time scale as

$$\tau_{f,\text{ind}} \sim \frac{R}{\sqrt{(V_\infty \, (1+a))^2 + (\Omega r \, (1+a'))^2}} \, , \tag{4.11}$$

where the term $V_\infty \, (1+a)$ is related to the actuator disk behaviour, whereas $\Omega r \, (1+a')$ is related to the impulsive effect of a discrete blade passage. This means that the maximum time scale is obtained through a continuous loading of the fluid, and the associated time scale is $R/(V_\infty \, (1+a))$; conversely, the passage of an isolated blade produces the minimum time scale described by equation (4.11). To complete the description, we consider non-dimensional correction factors $K_\square$ that account for the discrete nature of the blades and the rotor local solidity, to approximate how these factors affect the time scale of induction buildup. First, a blade with high solidity has a more distributed buildup of induction, behaving more similarly to an actuator disk. Thus, we define $K_\sigma = (1 - \sigma_r)$ where $\sigma_r$ is the blade solidity at the radial location $r$. Second, the impulsiveness is maximum at the blade tip, and diminishes at smaller radial locations, since the influence of the other blades increases closer to the symmetry axis. Thus, we define $K_r = r/R$. Both factors limit the impulsiveness and thus multiply the rotational velocity. So, we can compute the fluid time scale as

$$\tau_{f,\text{ind}} = \frac{R}{\sqrt{(V_\infty \, (1+a))^2 + (\Omega r \, (1+a') \, (K_\sigma K_r))^2}} \, , \tag{4.12}$$

or as

$$\tau_{f,\text{ind}} = \frac{R}{V_\infty} \frac{1}{\sqrt{(1+a)^2 + \lambda_r^2 (1+a')^2 (K_\sigma K_r)^2}} \tag{4.13}$$

after defining the local speed ratio of the blade section $\lambda_r = \Omega r / V_\infty$. In the test cases under analysis, the effect of $K_\sigma$ is almost negligible, whereas the effect of $K_r$ increases moving away from the tip. We note that analogous results might be obtained by defining the time scale differently, e.g., imposing $\tau_{f,\text{ind}} = 0.5 \, T/N_b/K_r$, where $T$ is the rotation period and $N_b$ the number of blades.

To summarize, the following system of four equations in four unknowns needs to be solved:

$$
\begin{cases}
Re_{\text{ind}} = \dfrac{\rho_f \left|V_{f,\text{ind}} - V_{p,\text{ind}}\right| d_p}{\mu_f} & \text{equation (4.6)} \\[2ex]
\psi_{\text{ind}} = 3 \, \dfrac{k \, Re_{\text{ind}}^{1/3} - \tan^{-1}\left(k \, Re_{\text{ind}}^{1/3}\right)}{k^3 \, Re_{\text{ind}}} & \text{equation (4.8)} \\[2ex]
Stk_{\text{ind}} = \dfrac{\tau_{p,0} \, \psi_{\text{ind}}}{\tau_{f,\text{ind}}} & \text{equation (4.7)} \\[2ex]
V_{p,\text{ind}} = \dfrac{V_{f,\text{ind}}}{1 + Stk_{\text{ind}}} & \text{equation (4.5)}
\end{cases} \tag{4.14}
$$

where the four unknowns are highlighted on the left-hand side, whereas $V_{f,\text{ind}}$ is known from equation (4.1), $\tau_{f,\text{ind}}$ is known from equation (4.12), and $\tau_{p,0} = (\rho_p d_p^2)/(18\mu_f)$.

The system can be conveniently solved iteratively using a fixed-point iteration method. The induced velocity of the particle at $t_p = 0$, i.e., at the blade section, is initialised as $V_{p,\text{ind}} = 0$. Then, the equations are solved sequentially:

1. $Re_{\text{ind}}$ is computed with equation (4.6);
2. $\psi_{\text{ind}}$ is computed through equation (4.8);
3. $Stk_{\text{ind}}$ is computed through equation (4.7);
4. $V_{p,\text{ind}}$ is updated with equation (4.5).

The loop is repeated from 1. to 4. until an equilibrium state is found; we monitor convergence through the difference of $Re_{\text{ind}}$ in two successive iterations. The minimum value of $Re_{\text{ind}}$ is set to twice the machine epsilon to avoid division by zero. The method was found to converge in less than 20 iterations to a tolerance of $10^{-6}$ in the cases under analysis; the rate of convergence increases with decreasing $Stk_{\text{ind}}$.

Once the magnitude of the induced velocity of the particle is known, its direction is determined from the induced velocity of the carrying phase:

$$
\boldsymbol{V}_{p,\text{ind}} = V_{p,\text{ind}} \frac{\boldsymbol{V}_{f,\text{ind}}}{||\boldsymbol{V}_{f,\text{ind}}||_2} \, , \tag{4.15}
$$

and the relative velocity between the particle and the section is given by

$$
\boldsymbol{V}_{p,\text{rel}} = \boldsymbol{V}_\infty + \boldsymbol{V}_{p,\text{ind}} - \boldsymbol{\Omega} \times \boldsymbol{r} \, . \tag{4.16}
$$

The angle of attack of a particle with respect to a section, $\alpha_{\text{part}}$, is then defined as the angle between $\boldsymbol{V}_{p,\text{rel}}$ and the chord of the section.

### 4.1. *Scale separation between the section and the induction zone*

The model provides a prediction of a particle's delayed response to the induced velocity field generated by a rotor. To validate the model, we recover once more to the comparison of 2D simulations with the 3D reference solutions computed previously. In 2D, one might want to initialise a parcel with its velocity vector in the aerodynamic sectional flow-field, i.e., in non-equilibrium; however, results would then depend on the release location.

A straightforward application is possible if there is a scale separation between the induction and section Stokes numbers, such that $Stk_{\text{ind}} \ll Stk_{\text{sec}}$. This occurs for $R \gg c$. If this is verified, it is possible to use the particle angle of attack $\alpha_{\text{part}}$ as the angle of attack of the carrying phase. Indeed, $Stk_{\text{ind}} \ll Stk_{\text{sec}}$ implies that:

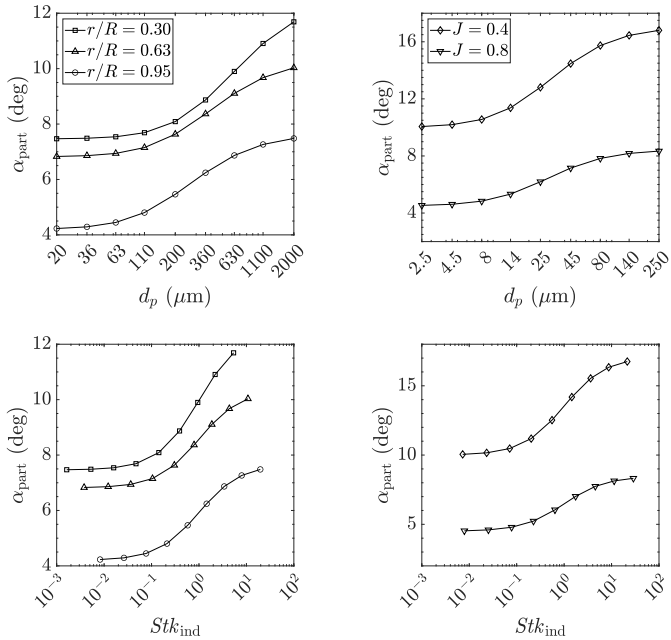

Figure 13. Particle angle of attack distribution according to their diameter (top) and induction Stokes number (bottom). Left: wind turbine blade sections. Right: propeller section at different advance ratios.

1. if $Stk_{ind} \ll 1$, the model outputs $V_{p,ind} \sim V_{f,ind}$, and the flow field around the section computed with the induced velocity of the particle will be similar to the actual flow field;

2. if $Stk_{ind} \sim 1$, then $Stk_{sec} \gg 1$; thus, the model outputs $V_{p,ind}/V_{f,ind} \sim 0.5$, and the the flow field computed using the induced velocity of the particle will differ from the actual flow field around the section; however, the particles' trajectories will be little affected by the airfoil itself leading to a locally ballistic solution;

3. if $Stk_{ind} \gg 1$, then $Stk_{sec} \gg Stk_{ind} \gg 1$; the model outputs $V_{p,ind} \sim 0$ and a globally ballistic solution, now independent from the induced velocity, is retrieved.

### 4.2. *Validation of the particle induction model*

All 2D simulations are recomputed according to the delay model described above and are identified with *2D Particle Ind*. The angles of attack $\alpha_{part}$ are reported in figure 13 for the different particle diameters. Since the fluid time scale is a function of the radial location, and so is $Stk_{ind}$, the response particles of the same size are different along the wind turbine blade. A milder dependency is seen on the freestream wind speed of the propeller. This also shows that the equilibrium condition with the induced flow of particles of poly-dispersed clouds depends on the particle diameter. Thus, a different condition should be imposed for each diameter of the particle distribution in a sectional simulation.

We recompute the section-normalised collection efficiency error according to equation (3.1). Results are reported in figure 14 for both the wind turbine blade and the propeller test cases. Previous results with *2D Ind* and *2D Geom* are included for comparison. This time, the error is shown as a function of $Stk_{ind}$. The model effectively reduces the error of the simulations in all the cases under analysis, and its application produces a collection efficiency almost equivalent to the 3D solution in the transition region. Moreover, we can analyse the evolution of the error of the limiting cases as a function of $Stk_{ind}$. With this new variable, the transition region from particles in equilibrium with the induced flow field to particles unaffected by it is located at $0.1 \lesssim Stk_{\text{ind}} \lesssim 10$. Notably, the maximum values of $Stk_{\text{ind}}$ (now accounting for the non-Stokesian drag by explicitly computing $\psi_{\text{ind}}$) for the cases shown previously in figure 11 are also smaller and aligned with this range: on the wind turbine section, when neglecting $\psi_{\text{ind}}$, $d_p = 630\,\mu\text{m}$ and $d_p = 2000\,\mu\text{m}$ led to $Stk_{\text{ind},0} \sim 10$ and $Stk_{\text{ind},0} \sim 100$, respectively; the values including $\psi_{\text{ind}}$ are $Stk_{\text{ind}} \sim 3$ and $Stk_{\text{ind}} \sim 20$. Similarly, on the propeller section, $d_p = 80\,\mu\text{m}$ and $d_p = 250\,\mu\text{m}$ now lead to $Stk_{\text{ind}} \sim 4$ and $Stk_{\text{ind}} \sim 30$. Although better estimates of $Stk_{\text{ind}}$ are possible, results prove that the quantity is representative of the behaviour of particles in the induction field upstream of a rotor. Moreover, we have shown that simple modelling of the induced velocity field $v_{f,\text{ind}}$ and of particles' interaction with it can lead to satisfactory results.

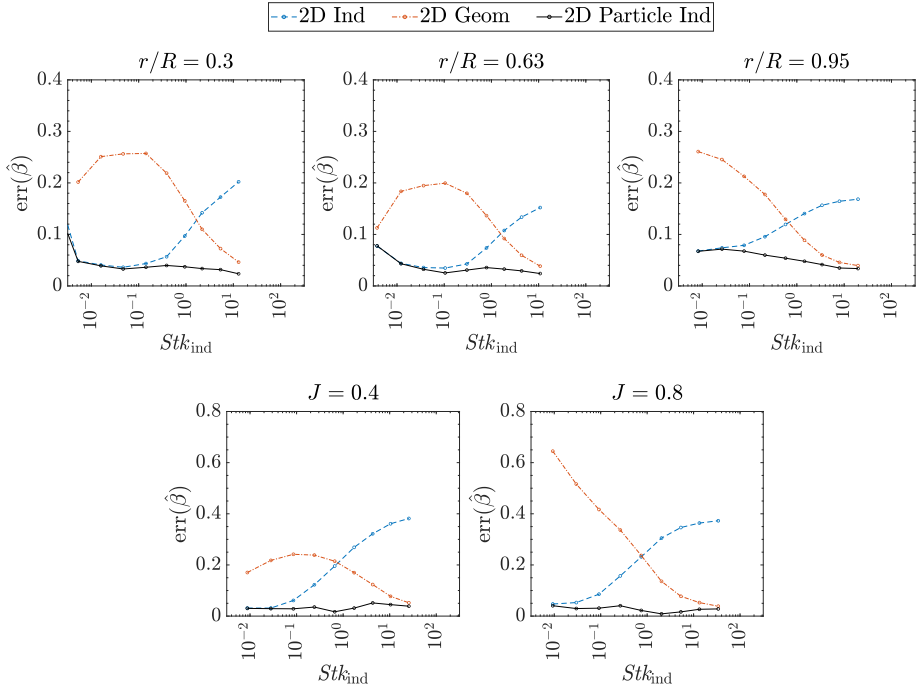

Figure 14. Error of 2D simulations with respect to the 3D solution, computed with equation (3.1), as a function of the induction Stokes number. Top: wind turbine sections ($r/R = 0.30$, $r/R = 0.63$, $r/R = 0.95$) at fixed operating conditions. Bottom: propeller section at $r/R = 0.80$ at different advance ratios ($J = 0.4$, $J = 0.8$). $d_p = 3600\,\mu\text{m}$ was included on the wind turbine section at $r/R = 0.30$ to reach $Stk_{\text{ind}} > 10$.

The results of the section-normalised collection efficiency are shown in figure 15 for a selection of representative cases in the transition region. On the wind turbine blade, the particle diameters are $d_p = 200\,\mu\text{m}$ and $d_p = 630\,\mu\text{m}$. The corresponding $Stk_{\text{ind}}$ on the three sections, from root to tip, is $\sim 0.15$, $\sim 0.30$, and $\sim 0.6$ for the first droplet size and $\sim 1$, $\sim 2$, and $\sim 4$ for the second. On the propeller section, the particle diameters are $d_p = 14\,\mu\text{m}$ and $d_p = 45\,\mu\text{m}$. The corresponding $Stk_{\text{ind}}$ is $\sim 0.25$ for the first particle size and $\sim 2$ for the second at both advance ratios. On the wind turbine blade, results are slightly off on the suction side ($s > 0$), whereas they are accurate on the pressure side ($s < 0$). Conversely, on the propeller, the results are particularly accurate on the suction side ($s > 0$), and slightly less on the pressure side ($s < 0$). The 2D ballistic solution for *2D Particle Ind* is also shown with a dotted line. The 3D results converge to the *2D Particle Ind* ballistic solution, similarly to what was observed with *2D Ind* and *2D Geom* and their respective ballistic limits in § 3. This shows that the particles become ballistic locally, around the section, while they are still in partial equilibrium with the induced velocity field. Particles reach the *2D Geom* ballistic solution once they become unaffected by the induced velocity field.

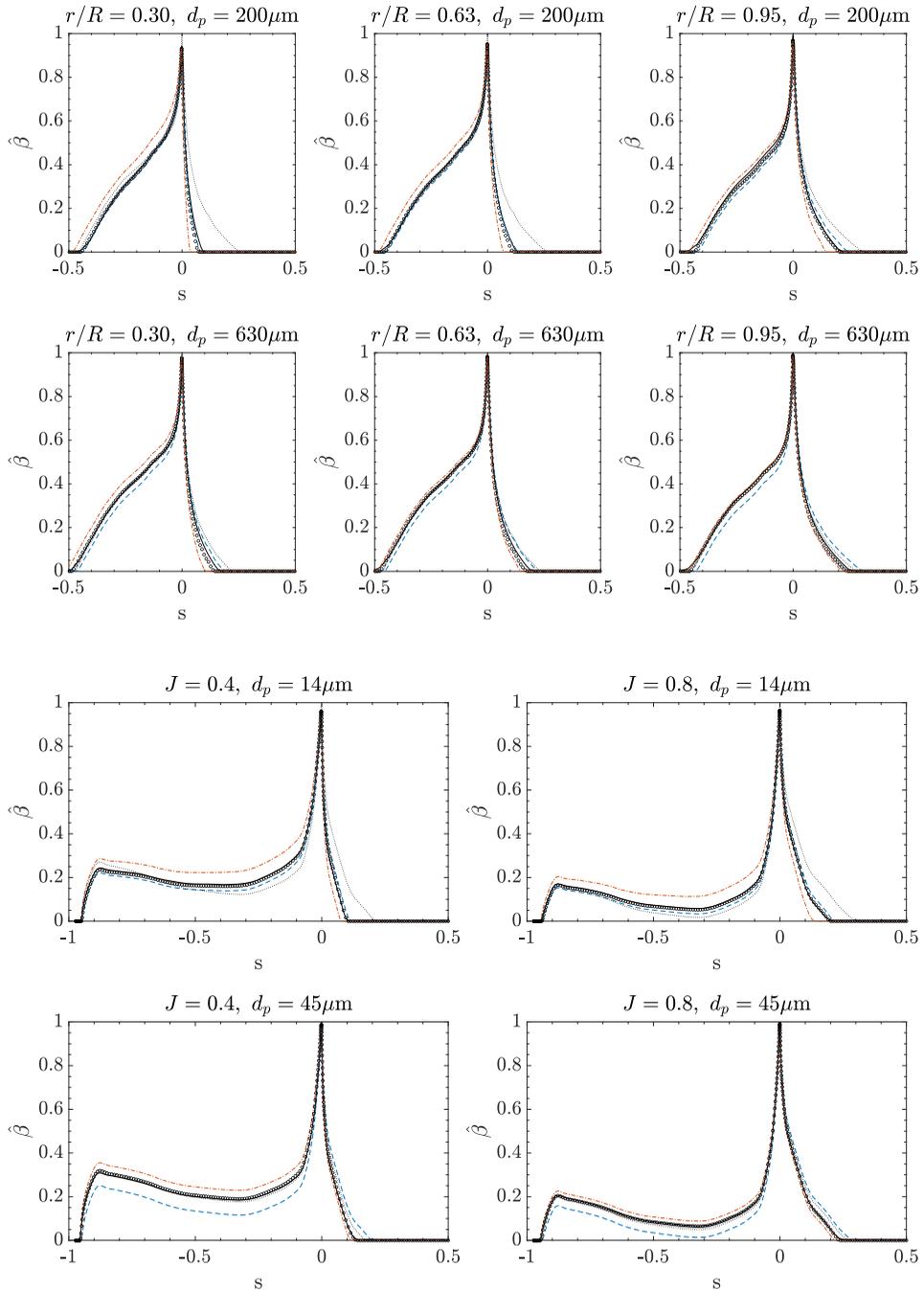

Figure 15. Section-normalised collection efficiency $\hat{\beta}$ as a function of the normalised curvilinear abscissa $s$ of the section. Results now include the 2D simulations with the model of particle behaviour in the induction field, identified with (∘). The ballistic limit of the solution at the section considering $\alpha_{\mathrm{part}}$ is shown with dotted lines (··). The first two rows show results on the wind turbine blade (each column identifies a radial location). The last two rows show results on the propeller section (each column identifies an advance ratio). A different droplet size is considered in each row.

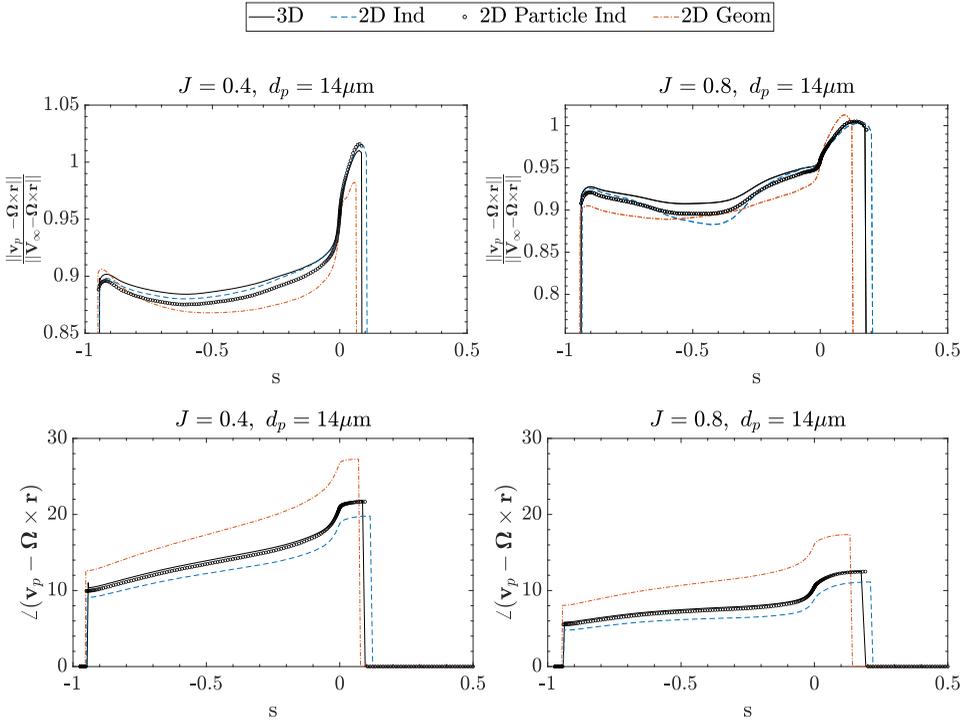

Figure 16. Particle-surface relative velocity at impact as a function of the normalised curvilinear abscissa $s$ of the propeller section at $r/R = 0.8$, including the results with the particle delay model. Each column represents a propeller advance ratio ($J = 0.4$, $J = 0.8$). Top: non-dimensionalised impact velocity magnitude. Bottom: angle of the in-plane velocity component with respect to the section chord.

Finally, the relative velocity vector of the particles with $d_p = 14\,\mu$m at impact on the propeller section is presented in figure 16 as magnitude and phase computed with respect to the chord of the section. The correct relative trajectory is retrieved, leading to a good prediction of the impact angle. Moreover, specific phenomena can be captured, such as the interaction of the particles with the section's boundary layer. This leads to a dip in the velocity magnitude at $J = 0.8$, which is overestimated with *2D Ind* and not captured with *2D Geom*. We also present the particle axial velocity field around the section of the wind turbine blade at $r/R = 0.95$ for $d_p = 200\,\mu$m computed with the different methods in figure 17. The particle trajectories are superimposed, highlighting the limiting trajectories. The freestream velocity set in the *2D Ind* and *2D Geom* cases is wrong, whereas the *2D Particle Ind* approach provides a better approximation. Both the particle velocity field and the trajectories around the section are better captured with the *2D Particle Ind* approach; differences in the particle velocity field around the section arise since the carrying field around the section is different.

Overall, the results show good agreement in all the test cases. The simplicity of the model highlights that its parameters represent the key physical mechanisms governing the transition region. Its validation through 2D simulations demonstrates that these mechanisms are correctly captured, and the model outputs the response of a particle to the induced velocity field. Moreover, the validation showed that accurate 2D simulations of particle impact are achievable for any droplet and rotor size with the *2D Particle Ind* approach when $Stk_{\text{sec}} \gg Stk_{\text{ind}}$.

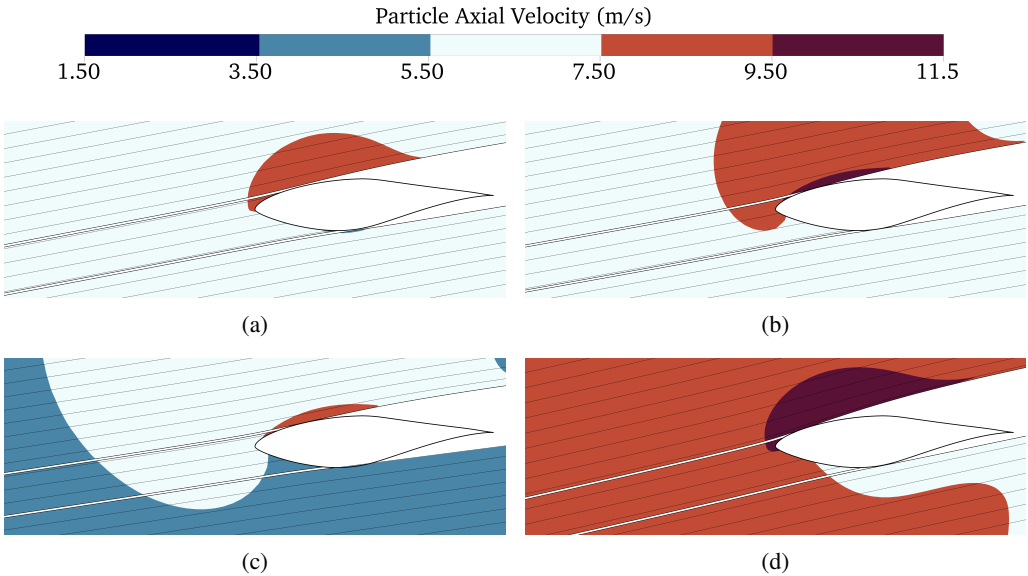

Figure 17. Visualization of the axial component of the particle velocity field together with the particles' trajectories, obtained with different simulation methods. The limiting trajectories are also highlighted. Wind turbine rotor, $r/R = 0.95$, $d_p = 200\,\mu\text{m}$. (a) 3D solution; (b) *2D Particle Ind* solution; (c) *2D Ind* solution; (d) *2D Geom* solution.

## 5. Particle transport regimes and relation with rotor size

Given all the results shown, we can identify the following regimes of particle transport in axial rotor induced flow fields, schematically represented in figure 18:

1. for $Stk_{\text{sec}} \lesssim 10^{-2}$, either particles don't impact the section, or second order effects become relevant. In this case, sectional simulations can be inaccurate; however, the impinging mass is small with respect to the section, so the dimensional error is small;

2. for $Stk_{\text{sec}} \gtrsim 10^{-1}$ and $Stk_{\text{ind}} \lesssim 10^{-1}$, particles are in equilibrium with the induced velocity field and filter out second order effects in the section flow field; thus, particles reach the section with $\alpha_{\text{aero}}$, and sectional simulations are accurate if the external velocity of the section, i.e. the pressure distribution on the section, is correct. In this regime, the *2D Ind* approach can be applied directly;

3. for $10^{-1} \lesssim Stk_{\text{ind}} \lesssim 10^{1}$, particles transition from the equilibrium with the rotor-induced velocity field to being unaffected by it. If $Stk_{\text{ind}} \ll Stk_{\text{sec}}$, the *2D Particle Ind* approach can be used to compute the sectional flow field without introducing significant error when computing the particle deposition;

4. for $Stk_{\text{sec}} \gtrsim 10^{1}$, the solution is independent of the flow field around the section, and becomes locally ballistic; however, the solution can depend on the induced velocity component of the particle;

5. for $Stk_{\text{ind}} \gtrsim 10^{1}$, particles don't respond to the induced velocity field, and if $Stk_{\text{sec}} \gtrsim 10^{1}$, the global ballistic solution on the surface, represented by the *2D Geom* case, is reached.

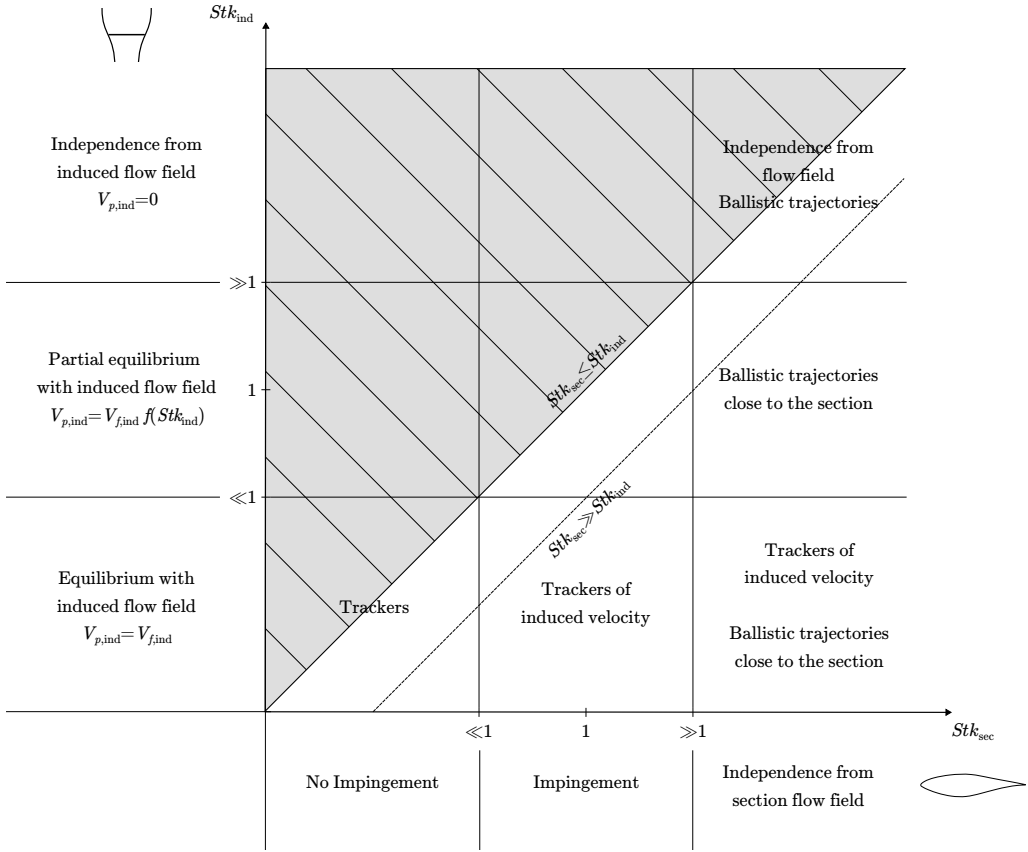

Figure 18. Schematic representation of the regimes encountered by particles immersed in the flow field of an isolated section, immersed in the induced flow field, and the resulting regimes on a rotor section.

It is important to remark that clouds are generally poly-dispersed, and the phenomena observed vary according to the radial position. Thus, different equilibrium conditions might coexist at a fixed radial location for different particle diameters, and at different radial locations for a fixed particle size. Moreover, we remark that we neglected particle breakup. A large droplet might break due to the effect of aerodynamic forces or after splashing on the surface. If this occurs close to the surface, the sectional Stokes number of the secondary droplets would be much lower, and they would need to be tracked on the correct sectional flow field. We have shown that the solution obtained on large droplets using *2D Particle Ind* is correct under the assumption $Stk_{ind} \ll Stk_{sec}$, so we infer that it might provide the correct initial conditions to the secondary droplets, to be tracked in the correct local flow field.

It is interesting to relate $Stk_{ind}$ with the size of a rotor and its rotational velocity to estimate where the transition regime is located. We consider the tip section of a generic low-solidity rotor and use the estimate of equation 4.11 for the fluid time scale and water droplets for the particle time scale. The tip is where most of the mass is collected and thus the region most subject to degradation. We provide two estimates in figure 19. On the left, we present the combination of droplet diameter, rotor diameter, and $V_{rel}$ (the fluid velocity seen by the section) that leads to $Stk_{ind,0} = 1$. For a high-solidity rotor, the left axis would likely be related more to the freestream wind speed than the tip speed, although this should be verified. On the right, this is further simplified assuming $\Omega R \gg V_\infty$, so that $V_{\mathrm{rel}} \sim \Omega R$. In this case, the fluid time scale of induction can be simplified to $\tau_{f,\mathrm{ind}} \sim 1/\Omega$, and different values of $Stk_{\mathrm{ind},0}$ can be computed for a combination of droplet diameters and rotational speeds. It is worth remarking that the actual values of $Stk_{\mathrm{ind}}$ are similar to those identified by the continuous lines on the plot for small $Stk_{\mathrm{ind},0}$ only. For increasing $Stk_{\mathrm{ind},0}$, $Re_p$ increases and and $Stk_{\mathrm{ind}} < Stk_{\mathrm{ind},0}$. Thus, the actual limiting values of the transition region are found with larger particles.

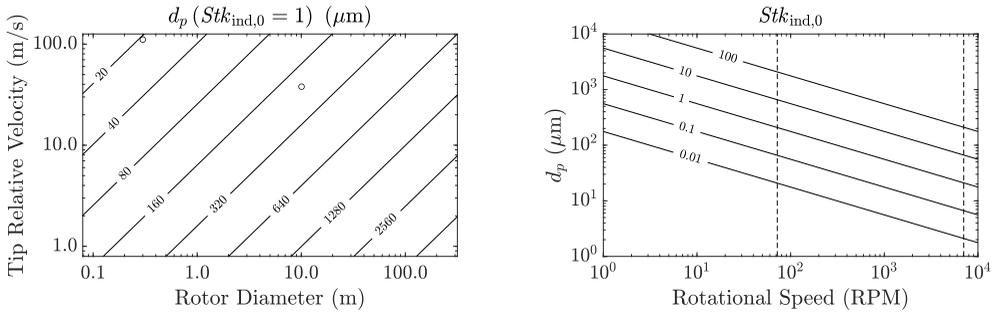

Figure 19. Location of the transition region from particles in equilibrium with the induced flow field to particles unaffected by it, considering water droplets and the tip section of a low-solidity rotor. Left: isolines of diameters $d_p$ in μm leading to $Stk_{\mathrm{ind},0} = 1$ as a function of the rotor diameter and the velocity seen by the tip of the blade. The circle markers (∘) show the parameters used in this work. Right: isolines of $Stk_{\mathrm{ind},0}$ as a function of the rotational speed and the droplet diameter, considering a large tip-speed ratio ($\Omega R \gg V_\infty$). The dashed lines (−−) show the rotational speeds used in this work.

For fast rotating rotors ($\Omega \gtrsim 1000\,\mathrm{rpm}$) the transition region is centred at $d_p \lesssim 50\,\mu\mathrm{m}$. The range of droplets is crucial for ice accretion, and the rotor speed is typical of small drones, aircraft propellers, distributed-electric propulsion and urban air mobility applications. For $100\,\mathrm{rpm} \lesssim \Omega \lesssim 1000\,\mathrm{rpm}$, the transition region is centred between $d_p \sim 50\,\mu\mathrm{m}$ and $d_p \sim 200\,\mu\mathrm{m}$. This range is relevant for in-flight icing in freezing drizzle conditions, with droplets up to $\sim 500\,\mu\mathrm{m}$ found for $Stk_{\mathrm{ind},0} \lesssim 10$. Rotors operating in this regime include the main helicopter rotors, low-noise operating propellers for vertical take-off and landing aircraft, and small wind turbines with diameters up to $\sim 5$ meters. For $10\,\mathrm{rpm} \lesssim \Omega \lesssim 100\,\mathrm{rpm}$, the transition region is centred between $d_p \sim 200\,\mu\mathrm{m}$ and $d_p \sim 500\,\mu\mathrm{m}$, with $Stk_{\mathrm{ind},0} \lesssim 10$ found for droplets up to $d_p \sim 2000\,\mu\mathrm{m}$. This range is relevant for freezing rain conditions and rain erosion. Rotors operating in this regime include onshore and offshore wind turbines with diameters up to $\sim 150\,\mathrm{m}$. Larger rotors typically have $\Omega \lesssim 10\,\mathrm{rpm}$ at rated wind speed, and the transition region is still relevant for freezing rain and erosion. This large design space still needs to be explored exhaustively. Estimates with sand, an important cause of erosion and whose density is of the same order of magnitude as water, are similar.

It is finally worth mentioning the problem of rotor scaling in experiments involving multi-phase flows, in view of the regimes identified in this work. Given the scaled rotor and its scaled operating conditions, if the particle size is chosen to match the Stokes number at the section $Stk_{\mathrm{sec}}$, it is not guaranteed, in general, that the Stokes number will also match in the stream tube upstream. In other terms, by denoting with $\square_{\mathrm{FS}}$ the full-scale condition and with $\square_{\mathrm{S}}$ the scaled condition, $Stk_{\mathrm{sec,FS}}(r) = Stk_{\mathrm{sec,S}}(r)$ does not imply $Stk_{\mathrm{ind,FS}}(r) = Stk_{\mathrm{ind,S}}(r)$. Moreover, if the condition on the induced flow field is matched, e.g., at the tip, this does not imply that the condition is satisfied along the entire blade span, and the particles in the scaled conditions might reach a section with an angle of attack different from that of the full-scale condition.

## 6. Conclusions

In this work, we have studied numerically the behaviour of particles immersed in rotor-induced flow fields and impacting the surface of rotor blades. We considered a wind turbine rotor and a propeller operating in a uniform freestream flow field aligned with the rotor blade rotation axis. To obtain a steady state solution, we solved either the 3D flow field in a rotating reference frame or 2D sectional flow field in the section reference frame. Individual particles in initial equilibrium with the carrying phase were tracked with a Lagrangian approach under one-way coupling assumptions. The section-normalised collection efficiency was the main parameter used to analyse particle behaviour.

We first considered the 2D solutions obtained by including or neglecting the aerodynamic induced velocity vector in the sectional computations. These represent the standard approach used in the literature and were found to represent two limiting solutions that can be characterised with the classical sectional Stokes number and by introducing an induction Stokes number. The 2D limiting solution for small particles is obtained for $Stk_{ind} \lesssim 0.1$, and 2D simulations require as necessary condition that the external velocity is correct. For large particles with $Stk_{ind} \gtrsim 10$ and $Stk_{sec} \gtrsim 10$, it is sufficient that the geometric orientation of the section is correct.

In specific conditions, the 3D solution was found to lie between the limiting solutions. Indeed, the analysis of the error between 2D and 3D solutions highlighted the existence of a transition regime from particles in equilibrium with the induced flow field to particles unaffected by it. This regime corresponds to an induction Stokes number $0.1 \lesssim Stk_{ind} \lesssim 10$. This also means that classical approaches for sectional simulations can introduce a systematic error in the solution. To overcome this limitation, we proposed a model for the delayed response of a particle to the induced velocity field.

The delay model represents the induced velocity field sampled by the particle with an exponential function. The solution at impact on the rotor blade led to a simple solution for the induced velocity component of the particle velocity, expressed as a function of $Stk_{ind}$ only. This is computed by considering the particle Reynolds number $Re_{ind}$ to compute the particle time scale in the ultra-Stokesian regime and modelling the fluid time scale. The latter considers the rotor radius as the relevant length scale and a modified section velocity to account for the discrete or continuous buildup of induction at each blade passage.

The model was successfully validated on the test cases assuming a scale separation between the sectional and induction Stokes numbers ($Stk_{ind} \ll Stk_{sec}$, which occurs for $c \ll R$). The assumption implies that the angle of attack of the particle computed from the model output can be applied to the carrying phase in 2D to then compute particle impingement on the section. Although more accurate estimates of $Stk_{ind}$ or the induced velocity field are possible, the results proved that: (i) the local distribution of the impinging volume of particles is a function of their behaviour in the stream tube upstream of the rotor disk, where induction is built; (ii) the behaviour of particles upstream can be represented with an induction Stokes number $Stk_{ind}$, such that particles are in equilibrium with the induced velocity field for $Stk_{ind} \lesssim 0.1$ and are unaffected by it for $Stk_{ind} \gtrsim 10$; (iii) upstream conditions can be modelled with a simple model, that relies on a representative, continuous velocity field and the computation of $Re_{ind}$ and $Stk_{ind}$ at the rotor disk. We have also shown that for $Stk_{sec} \gg Stk_{ind}$ the trajectories become insensitive to the section velocity field while being still affected by the induced velocity field, producing a ballistic behaviour only locally, around the section.

Since real clouds are usually polydispersed, each particle size would respond differently to the rotor-induced velocity field. This makes the use of delay models of particles fundamental in 2D simulations. Further validation will help identify the limits of the model proposed, possible extensions, as well as alternative methods to study and determine the delay of particles in the induction region in the stream tube upstream of a rotor. Future work might also be devoted to assessing scaling conditions for rotors in multi-phase flows.

**Declaration of interests** The authors report no conflict of interest.

**Author ORCIDs** F. Caccia, https://orcid.org/0000-0003-1331-3455; A. Guardone, https://orcid.org/0000-0001-6432-2461

## Appendix A. Lagrangian tracking algorithm in a rotating reference frame

We focus on the algorithm for the solution of equation (2.3) in a rotating reference frame. We developed an approach which avoids integrating the non-inertial forces. The velocity of each particle $\boldsymbol{v}_p$ is maintained in the same reference frame of the fluid velocity $\boldsymbol{v}_f$ according to the absolute velocity formulation as presented in Sec. 2.1, i.e., the instantaneous reference frame of the blade not accounting for the velocity induced by the frame motion. Thus, the integration of the acceleration of the particle is straightforward. E.g., we can use a forward Euler integration scheme with a time step $\Delta t$ to get the solution at the time-step $n + 1$ given the solution at the time-step $n$:

$$_n\boldsymbol{a}_p^{n+1} = f(_n\boldsymbol{v}_p^n - _n\boldsymbol{v}_f^n), \tag{A1}$$

$$_n\boldsymbol{v}_p^{n+1} = _n\boldsymbol{v}_p^n + _n\boldsymbol{a}_p^{n+1}\Delta t, \tag{A2}$$

$$_n\boldsymbol{x}_p^{n+1} = _n\boldsymbol{x}_p^n + _n\boldsymbol{v}_p^{n+1}\Delta t. \tag{A3}$$

where $\boldsymbol{a}_p$ denotes the acceleration of the parcel and $\boldsymbol{x}_p$ its position, whereas $_n\square^{n+1}$ denotes the solution at the time-step $n + 1$ expressed in the reference frame at time $n$. Then, since the reference frame rotates by a rotation vector $\Delta\boldsymbol{\theta} = \boldsymbol{\Omega}\Delta t$ with respect to the inertial frame between the time step $n$ and $n + 1$, the position vector of the particle is expressed in the new reference frame (the one at time $n + 1$), and so is the velocity vector:

$$_{n+1}\boldsymbol{v}_p^{n+1} = \mathcal{R}(_n\boldsymbol{v}_p^{n+1}, \Delta\boldsymbol{\theta}), \tag{A4}$$

$$_{n+1}\boldsymbol{x}_p^{n+1} = \mathcal{R}(_n\boldsymbol{x}_p^{n+1}, \Delta\boldsymbol{\theta}), \tag{A5}$$

where $\mathcal{R}(\boldsymbol{a}, \boldsymbol{b})$ is a function that rotates the reference frame of vector $\boldsymbol{a}$ by the rotation vector $\boldsymbol{b}$. In this case, the Rodrigues rotation formula is used. This will provide the new position of the particle in the stationary mesh, i.e., in the non-inertial frame. The same applies to the velocities. Thus, after locating the new position of the particle in the mesh or their intersection with a solid boundary, the integration can continue from equation A1. In this way, there is no need to integrate the non-inertial accelerations, avoiding the discretization error related to those terms.

### A.1. *Verification*

A cloud of Stokesian particles is immersed in a flow field with a prescribed velocity

$$\boldsymbol{v}_f(r) = r\hat{\boldsymbol{r}} \tag{A6}$$

assigned to the nodes of the computational grid, where $r$ is the radial distance from the rotation axis and $\hat{\boldsymbol{r}}$ is the unit radial vector. Particles are in initial equilibrium with the surrounding flow, i.e., $\boldsymbol{v}_p(\boldsymbol{x}, t = 0) = \boldsymbol{v}_f(\boldsymbol{x}, t = 0)$. The analytical solution is

$$\tilde{r}_p(t) = \left(\frac{\lambda_2 - 1}{\lambda_2 - \lambda_1}e^{\lambda_1 t} - \frac{\lambda_1 - 1}{\lambda_2 - \lambda_1}e^{\lambda_2 t}\right)\tilde{r}_p(0), \tag{A7}$$

where $\lambda_1$ and $\lambda_2$ are the eigenvalues of the system. The solution is computed in the inertial and non-inertial reference frames, the latter rotating with a constant angular velocity $\Omega = 2\pi$, for $N_p$ particles. The non-dimensional position error $\epsilon_j = (r_{p,j} - \tilde{r}_{p,j})/\tilde{r}_{p,j}$ is evaluated for $j \in [1, N_p]$, $\tilde{r}_p(0) \neq 0$ to build the error vector $\boldsymbol{\epsilon} \in \mathbb{R}^{N_p}$. The non-dimensionalisation makes the error dependent on the integration timestep only, and not on the initial position, i.e., $\epsilon_j = ||\boldsymbol{\epsilon}||_2/N_p \ \forall j \in [1, N_p]$. The non-dimensional position error $\epsilon_j = ||\boldsymbol{\epsilon}||_2/N_p$ is evaluated at $t = 1$ s, i.e., after one revolution, and shown in Figure 20. The solution is advanced in time with a forward Euler integration scheme. The error computed in the inertial and non-inertial reference frames overlap and the expected order of convergence is retrieved. A solution obtained by integrating the non-inertial forces would have the same order of convergence but include an additional error due to the integration of these terms, and its magnitude would be case-dependent.

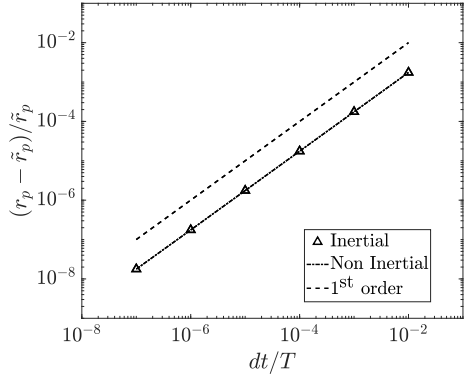

Figure 20. Normalized position error $(r_p - \tilde{r}_p)/\tilde{r}_p$ of a generic particle. The algorithm presented in Appendix A does not introduce an integration error due to non-inertial forces since these are not integrated. Thus, the errors computed in the inertial and non-inertial reference frames coincide.

| Number of points | $C_{F_{\text{inv}}}$ ($r = 0.30R$) | $C_{F_{\text{inv}}}$ ($r = 0.63R$) | $C_{F_{\text{inv}}}$ ($r = 0.95R$) |
|---|---|---|---|
| $24.1 \times 10^3$ | 0.968 | 1.21 | 1.26 |
| $\mathbf{48.9 \times 10^3}$ | 0.966 | 1.21 | 1.28 |
| $94.7 \times 10^3$ | 0.966 | 1.21 | 1.29 |

Table 8. 2D grid convergence analysis. The local relative velocity vector without induction is imposed at each section without transition modelling. $C_{F_{\text{inv}}}(r)$ is the inviscid component of the force coefficient at radial coordinate $r$, i.e., the integral of the pressure coefficient normalised by the local chord. The grid used for the simulations is highlighted in bold.

## Appendix B. Domain discretisation and grid independence

Grid convergence is evaluated on the inviscid force coefficient $C_{F_{\text{inv}}}$ computed on a set of three grids. Each grid is built considering a linear refinement factor of 1.4 in every direction.

### B.1. *Wind turbine rotor*

Grid convergence in 2D is evaluated by computing the flow field without induced velocities. The results are presented in table 8. After evaluating grid independence, the 2D unstructured grid used for computing the sectional flow field and the collection efficiency consists in $50 \times 10^3$ points, with 316 points on the airfoil, a far-field located at $100c$ distance, a first cell height in the anisotropic cells of the boundary layer of $10^{-6}c$ to ensure $y^+ < 1$ at the first layer, and a growth rate of 1.1.

For full 3D simulations, a cylindrical unstructured domain with a far-field distance of $50R$ is used. The results of the grid convergence analysis are presented in table 9. The grid with $6.6 \times 10^6$ points has a sufficient resolution and is chosen for the study. The blade surface is discretised with a triangulated structured grid of $163 \times 198$ points (chord × span). The chordwise discretisation of the airfoil has elements at the leading and trailing edge of $\frac{c}{1000}$ length, with $c$ being the chord of the section, and a maximum length of $\frac{c}{25}$. The maximum spanwise length is $\frac{R-R_0}{75}$, with $R_0 = 0.5083$ m denoting the root radial position. The first cell height is set to $2 \times 10^{-6}$m to ensure $y^+ < 1$ at the first layer on the whole blade, with a growth rate of 1.1. The farfield surface cell characteristic length is $10R$.

### B.2. *Small-scale propeller*

Grid convergence in 2D is evaluated by computing the flow field without induced velocities. The results are presented in table 10. The 2D unstructured grid used for computing the sectional flow-field and the collection efficiency consisted in $35 \times 10^3$ points, with 314 points on the airfoil, a farfield located at $100c$ distance with cells of $4c$ size, and a first cell height of $10^{-4}c$ to ensure $y^+ < 1$ at the first layer, and a growth rate of 1.1. For 3D simulations, a spherical domain of $5.6 \times 10^6$ points with a far-field distance of $100R$ was used. The results of grid convergence analysis are presented in table 11. The blade surface is discretised with a triangulated structured grid of $156 \times 140$ points (chord × span). The first cell height is set to $10^{-6}$m to ensure $y^+ < 1$ at the first layer on the whole blade, and the growth rate was is to 1.1.

| Number of points | $C_{F_{inv}}$ ($r = 0.30R$) | $C_{F_{inv}}$ ($r = 0.63R$) | $C_{F_{inv}}$ ($r = 0.95R$) |
|---|---|---|---|
| $2.5 \times 10^6$ | 0.788 | 0.831 | 0.523 |
| **$6.6 \times 10^6$** | 0.829 | 0.847 | 0.533 |
| $18.2 \times 10^6$ | 0.831 | 0.849 | 0.530 |

Table 9. 3D grid convergence analysis. Simulations are carried out at $7\,\mathrm{m\,s^{-1}}$ without transition modelling. $C_{F_{inv}}(r)$ is the inviscid component of the force coefficient at radial coordinate $r$, i.e., the integral of the pressure coefficient normalized by the local chord. The grid used for the simulations is highlighted in bold.

| Number of points | $C_{F_{inv}}$ ($r = 0.80R$) |
|---|---|
| $17.4 \times 10^3$ | 1.25 |
| **$33.8 \times 10^3$** | 1.24 |
| $67.5 \times 10^3$ | 1.24 |

Table 10. 2D grid convergence analysis. The local relative velocity vector without induction is imposed at each section without transition modelling. $C_{F_{inv}}(r)$ is the inviscid component of the force coefficient at radial coordinate $r$, i.e., the integral of the pressure coefficient. The grid used for the simulations is highlighted in bold.

| Number of points | $C_{F_{inv}}$ ($r = 0.80R$) |
|---|---|
| $2.1 \times 10^6$ | 0.902 |
| **$5.2 \times 10^6$** | 0.740 |
| $17.8 \times 10^6$ | 0.728 |

Table 11. 3D grid convergence analysis. Simulations are carried out at $J = 0.8$ without transition modelling. is imposed at each section. $C_{F_{inv}}(r)$ is the inviscid component of the force coefficient at radial coordinate $r$, i.e., the integral of the pressure coefficient. The grid used for the simulations is highlighted in bold.